\documentclass[letterpaper,10pt,twoside]{article}

\usepackage[utf8]{inputenc}  
\usepackage[english]{babel} 

\usepackage{lmodern}
\usepackage{amssymb,amsthm,mathtools} 
\usepackage{hyperref} 
\usepackage{geometry} 

\usepackage{xcolor,graphicx}
\graphicspath{{images/}}

\usepackage{pgfplots}
\pgfplotsset{compat=1.18}
\usepgfmodule{plot}
\usetikzlibrary{patterns}
\usetikzlibrary{calc}
\usetikzlibrary{decorations}
\usetikzlibrary{decorations.pathreplacing}
\usetikzlibrary{decorations.markings}

\newcommand{\im}{\mathrm{i}}

\newcommand{\R}{\mathbb{R}}
\newcommand{\bC}{\mathbb{C}}
\newcommand{\C}{\mathbb{C}}
\newcommand{\defeq}{\coloneqq}
\newcommand{\tens}{\otimes}
\DeclareMathOperator{\id}{id}
\newcommand{\xd}{\mathrm{d}}
\newcommand{\cH}{\mathcal{H}}
\newcommand{\po}{\mathsf{P}}
\DeclareMathOperator{\tr}{\mathrm{tr}}
\newcommand{\coh}{\mathsf{K}}

\newcommand{\Lr}{\mathrm{R}}
\newcommand{\Ll}{\mathrm{L}}
\newcommand{\Lcr}{\overline{\mathrm{R}}}
\newcommand{\Lcl}{\overline{\mathrm{L}}}
\newcommand{\Le}{\mathrm{e}}
\newcommand{\Lp}{\mathrm{p}}

\newcommand{\vac}{\mathbf{0}}
\newcommand{\tord}{\mathbf{T}}

\allowdisplaybreaks

\begin{document}


\begin{titlepage}
\title{\textbf{A time operator and the time-of-arrival problem in quantum field theory}}
\author{%
  Daniele Colosi\footnote{email: dcolosi@enesmorelia.unam.mx}\\
  Escuela Nacional de Estudios Superiores, Unidad Morelia,\\
  Universidad Nacional Autónoma de México,\\
  C.P.~58190, Morelia, Michoacán, Mexico
  \and Robert Oeckl\footnote{email: robert@matmor.unam.mx}\\
  Centro de Ciencias Matemáticas,\\
  Universidad Nacional Autónoma de México,\\
  C.P.~58190, Morelia, Michoacán, Mexico}
\date{UNAM-CCM-2024-1\\ 11 July 2024\\ 30 June 2025 (v2)}

\maketitle

\vspace{\stretch{1}}

\begin{abstract}

The Newton-Wigner states and operator are widely accepted to provide an adequate notion of spatial localization of a particle in quantum field theory on a spacelike hypersurface. Replacing the spacelike with a timelike hypersurface, we construct one-particle states of massive Klein-Gordon theory that are localized on the hypersurface in the temporal as well as two spatial directions. This addresses the longstanding problem of a "time operator" in quantum theory. It is made possible by recent advances in quantization on timelike hypersurfaces and the introduction of evanescent particles. As a first application of time-localized states, we consider the time-of-arrival problem. Our results are in accordance with semiclassical expectations of causal propagation of massless and massive particles. As in the Newton-Wigner case, localization is not perfect, but apparent superluminal propagation is exponentially suppressed.

\end{abstract}

\vspace{\stretch{1}}
\end{titlepage}



\section{Introduction}

It has long been a controversial subject in quantum theory whether a “time operator” could be constructed with features roughly analogous to the position operator of non-relativistic quantum mechanics. Famously, Wolfgang Pauli pointed out that the existence of such an operator would seem to conflict with the boundedness from below of the Hamiltonian \cite[p.~60]{Flu:encphys5-1}, see also \cite{Allcock:1969cq,MuLe:arrivaltime}. Of course, it should be remembered that this position operator does not arise from a spacetime context, but from the particular context of a canonical decomposition of the phase space of a system consisting of a single particle. Other systems do not in general have a natural “position” operator, even in non-relativistic quantum mechanics.\footnote{To begin with, many systems (as in quantum information for example) do not consist of particles with position degrees of freedom. Even if a system does consist exclusively of such particles, and even if there are position operators for the individual particles, there need not exist a sensible collective position operator.}

In any case, the core problem is that time is an external parameter in the standard formulation of quantum theory. In order to be able to treat time on an equal footing with space, we have to resort to a more powerful formulation of quantum theory. Here, we will use the Positive Formalism (PF) \cite{Oe:dmf,Oe:posfound}.
The basic idea is quite simple. In spacetime terms, particle localization at a given time $t$ refers to its position on the equal-time hypersurface at this time $t$. For a single particle without structure, this position constitutes half of its phase space, and gives rise upon quantization to a position operator. Replace now the standard equal-time hypersurface with a timelike hypersurface, say at a fixed coordinate value in a given spatial direction. The “position” of a particle within the hypersurface (and thus a canonical half of the phase space) now consists of the time coordinate in addition to the two tangential spatial coordinates. This seems to suggest that upon quantization we would indeed obtain a time operator (along with two spatial position operators). Unfortunately, for realistic dynamics it is not possible to isolate a single particle sector on a timelike hypersurface in a consistent quantum theory. Rather, even if we tried to start from a single-particle “quantum mechanical” setting, coherence of the theory automatically leads to a many-particle quantum field theoretic setting \cite{Oe:boundary}. Thus, we necessarily have to look for a “time operator” in the latter setting. This is what we shall do in the present work.

We consider in this work massive and massless Klein-Gordon theory in Minkowski space. After fixing conventions for the standard quantization on an equal-time hypersurface in Section~\ref{sec:slquant} we review in Section~\ref{sec:spaceop} the localized states of Newton and Wigner, as well as their localized measurement as a Positive Operator Valued Measure. In Section~\ref{sec:positionprob} we review the probability distribution of the spatial localization of an initially perfectly localized particle. We switch in Section~\ref{sec:Hstml} to our setting of interest with the quantization of Klein-Gordon theory on a timelike hyperplane at a fixed position in the $x_1$-coordinate. The first such quantization has been performed almost 20 years ago \cite{Oe:timelike,Oe:kgtl}, but was limited to the \emph{propagating sector} of the phase space. While this was sufficient for its application in asymptotic scattering theory \cite{CoOe:spsmatrix,CoOe:smatrixgbf}, it is not sufficient for our present purposes, where we also require the inclusion of the \emph{evanescent sector} of the phase space. 
This arises from the fact that in contrast to a spacelike hypersurface, the solutions of the Klein-Gordon equation in a neighborhood of a timelike hypersurface include evanescent waves in addition to propagating ones. This is a familiar circumstance in the case of the electromagnetic field, but applies to the scalar field as well.
The quantization of this evanescent sector was made possible only recently by the introduction of a novel twisted quantization scheme \cite{CoOe:locgenvac}. The complete quantization was performed in \cite{CoOe:evanescent}, giving rise to the concept of \emph{evanescent particle}. Based on this, we construct in Section~\ref{sec:timeop} time-localized states as well as the corresponding Positive Operator Valued Measure, including a time operator.

A second focus of the present work, in Section~\ref{sec:toa}, is the time-of-arrival problem, as a first application of time-localized states. For essentially the same reasons that prevent the notion of a time operator, the determination of the arrival time of a particle through a quantum measurement is much less straightforward in the standard formulation than the determination of the position at a given time. 
Such difficulty has led to numerous proposals in the literature, most of which are based on standard quantization techniques (in particular canonical quantization) within the framework of quantum mechanics, see e.g.\ \cite{MuLe:arrivaltime,Allcock:1969cq,Grot:1996xu,AnSa:time_2019}. It is interesting to notice that Kijowski's approach \cite{Kijowski:1974jx} exhibits some similarities to our proposal since he considers a timelike hyperplane and construct a probability density of passing through this hyperplane at a fixed time. However, the quantization scheme is still the standard one.

Using spatial transition amplitudes between time-localized states we extract detection probabilities as a function of arrival time. While there are many analogies to the Newton-Wigner setting, there is a crucial difference. While a particle localized on a spacelike hypersurface can only evolve to the future, a particle localized on a timelike hypersurface might move in either direction orthogonal to the hypersurface. This requires the introduction of an additional quantum number distinguishing the two possibilities.

We close the article with Section~\ref{sec:conclusions}, offering some conclusions and an outlook. Appendix~\ref{sec:app} collects some of the calculations underlying the results of Section~\ref{sec:toa}.


\section{Quantization on a spacelike hyperplane reviewed}
\label{sec:slquant}

Our quantization of massive Klein-Gordon theory on a spacelike hyperplane is standard. It is convenient to express this in the language of the instantaneous phase space $L$, its symplectic form $\omega$, complex structure $J$, and inner product $\{\cdot,\cdot\}$ \cite{Woo:geomquant}. For comparability, we follow \cite{CoOe:evanescent}. The complexified \emph{phase space} $L^\C$ at time $t$ is parametrized in terms of plane waves, with $E=\sqrt{k^2+m^2}$,
\begin{equation}
 \phi(t,x)=\int\frac{\xd^3 k}{(2\pi)^3 2E}
 \left(\phi^\text{a}(k) e^{-\im(E t-k x)}+\overline{\phi^\text{b}(k)} e^{\im(E t-k x)}\right) .
 \label{eq:kgmodes}
\end{equation}
The complexified \emph{symplectic form} $\omega:L^\C\times L^\C\to\C$ that partially encodes the classical dynamics is given by,
\begin{equation}
  \omega(\phi,\eta)
  =\frac12\int\xd^3 x\, \left(\eta(t,x)\partial_0 \phi(t,x)-\phi(t,x)\partial_0\eta(t,x)\right)
  =\frac{\im}{2}\int\frac{\xd^3 k}{(2\pi)^3 2E}
  \left(\eta^\text{a}(k)\overline{\phi^\text{b}(k)}-\phi^\text{a}(k)\overline{\eta^\text{b}(k)}\right) .
  \label{eq:sfkgm}
\end{equation}
The \emph{complex structure} given by the operator $J:L^\C\to L^\C$ encodes the Minkowski vacuum. Its eigenspaces are the spaces of \emph{positive} and \emph{negative energy} solutions with eigenvalues $-\im$ and $\im$ respectively. It is given by,
\begin{equation}
  (J(\phi))^{\text{a/b}}(k) =-\im \phi^{\text{a/b}}(k) .
\end{equation}
We obtain the \emph{complex inner product} on the phase space $L$,
\begin{equation}
  \{\phi,\eta\}\defeq
  4\im\omega(\overline{\phi^+},\eta^+)
  = 2\omega(\phi,J \eta)+2\im\omega(\phi,\eta)
  =2\int\frac{\xd^3 k}{(2\pi)^3 2E}
  \phi^\text{a}(k)\overline{\eta^\text{b}(k)} .
  \label{eq:rjip}
\end{equation}
Here, $\phi^+$ denotes the positive energy component of $\phi$, i.e., its projection onto the eigenspace with eigenvalue $-\im$ under $J$. Note that the second expression is the one most commonly found in the literature, e.g.\ \cite{BiDa:qftcurved}.
The inner product determines the \emph{commutation relations} between creation and annihilation operators,
\begin{equation}
  [a_{\eta},a_{\phi}^{\dagger}]=\{\phi,\eta\} .
  \label{eq:ccr}
\end{equation}
We use the following notation for $n$-particle states, where $\vac$ denotes the vacuum state,
\begin{equation}
    \Psi_{\xi_1,\ldots,\xi_n}\defeq a^\dagger_{\xi_1}\cdots a^\dagger_{\xi_n} \vac .
\end{equation}
In scattering problems it is common to label creation and annihilation operators by 3-momentum rather than by phase space elements. To connect to this more conventional point of view consider plane wave solutions, ($E_p$ is the energy for the 3-momentum $p$),
\begin{equation}
   \Phi_p(t,x)\defeq \frac{1}{\sqrt{2}}\left(e^{-\im (E_p - p x)} + e^{\im (E_p - p x)}\right) .
\end{equation}
The coefficients $(\Phi_p)^{a,b}(k)$ in the expansion (\ref{eq:kgmodes}) are then given by suitably normalized delta functions and the associated creation and annihilation operators recover the usual momentum space commutation relations,
\begin{equation}
  a_p\defeq a_{\Phi_{p}},\quad a^\dagger_p\defeq a^\dagger_{\Phi_{p}},\quad
  [a_{p'},a^\dagger_p]=\{\Phi_{p},\Phi_{p'}\}
  =(2\pi)^3 2 E_p \delta^3(p-p') .
\end{equation}
It is also instructive to consider the usual field operators $\hat{\phi}(t,x)$. These correspond to solutions that are plane waves in momentum space with coefficients,
\begin{equation}
  (\widetilde{\Phi}_x^t)^{\text{a,b}}(k)=\frac{1}{\sqrt{2}}e^{\im(E t-k x)}.
  \label{eq:posop}
\end{equation}
In this way, $\hat{\phi}(t,x)$ is the usual sum of a creation and annihilation operator,
\begin{equation}
  \hat{\phi}(t,x)=a_{\widetilde{\Phi}_x^t}+a^\dagger_{\widetilde{\Phi}_x^t},
\end{equation}
leading to the Feynman propagator,
\begin{multline}
  G_{\text{F}}(t,x;t',x') = \im\langle\vac,\tord\hat{\phi}(t',x')\hat{\phi}(t,x)\vac\rangle
  =\im\Theta(t'-t) [a_{\widetilde{\Phi}_{x'}^{t'}},a^\dagger_{\widetilde{\Phi}_x^t}]
  +\im\Theta(t-t') [a_{\widetilde{\Phi}_x^t},a^\dagger_{\widetilde{\Phi}_{x'}^{t'}}] \\
  =\im\Theta(t'-t) \{\widetilde{\Phi}_{x}^{t},\widetilde{\Phi}_{x'}^{t'}\}
  +\im\Theta(t-t') \{\widetilde{\Phi}_{x'}^{t'},\widetilde{\Phi}_x^t\}
  = \im \int \frac{\xd^3 k}{(2\pi)^3 2E} e^{-\im(E|t'-t| - k ( x'-x ))} .
  \label{eq:feynmanprop}
\end{multline}


\section{Localized states and the Newton-Wigner Operator}
\label{sec:spaceop}

There is a well established notion of a position operator in quantum field theory, the Newton-Wigner position operator \cite{NeWi:locstates}. In the present section we briefly review this operator and the associated position-localized states of Klein-Gordon theory. Our treatment builds on that of Wightman \cite{Wig:localizability}.

Fix a time $t\in\R$. The principal ingredient is the solution $\Phi_{x}^t\in L$ (recall the parametrization (\ref{eq:kgmodes})) with $x\in\R^3$ an element of position space, given by,
\begin{equation}
  (\Phi_{x}^t)^{\text{a,b}}(k)=\sqrt{E} e^{\im(E t- k x)} .
  \label{eq:locsol}
\end{equation}
This can be thought of as encoding a single particle at time $t$, localized at position $x$. (In the case $t=0$ this is essentially the expression following (9a) in \cite{NeWi:locstates}.) Note that compared to expression \eqref{eq:posop} there is a relative factor $\sqrt{2 E}$. Crucially, this leads to the inner product \eqref{eq:rjip} between these solutions being a delta function,
\begin{equation}
  \{\Phi_{x'}^t,\Phi_{x}^t\}=\delta^3(x-x') .
\end{equation}
This in turn leads to the \emph{completeness relation},
\begin{equation}
  \{\xi,\eta\}=\int\xd^3 x\, \{\xi,\Phi_{x}^t\} \{\Phi_{x}^t,\eta\}
  \qquad\forall\,\xi,\eta\in L .
\end{equation}
Upon quantization this yields a corresponding completeness relation on the 1-particle Hilbert subspace $\cH^1\subseteq \cH$ of the full Hilbert space of states. In terms of creation and annihilation operators we can write this as,
\begin{equation}
  \id^1=\int\xd^3 x\, a^\dagger_{t,x} \po_0 a_{t,x} .
\end{equation}
Here, $\po_0\defeq |\vac\rangle\langle\vac|$ is the projector onto the vacuum state and we use the notation $a_{t,y}\defeq a_{\Phi_{y}^t}$. We can now define a quantization map from real-valued functions on position space $\R^3$ to the space of self-adjoint operators on $\cH$,
\begin{equation}
  \hat{f}\defeq \int\xd^3 x\, f(x)\, a^\dagger_{t,x} \po_0 a_{t,x} .
\end{equation}
Since this operator maps $\cH^1$ to itself, and is positive if $f$ is non-negative this defines a \emph{positive operator valued measure} (POVM) on $\cH^1$. In particular, this makes possible an interpretation in terms of position measurements and their probabilities. Let $U\subseteq\R^3$ be a Lebesgue-measurable subset, then the quantization $\hat{\chi}_U$ of its characteristic function $\chi_U$ yields the projection operator that encodes measurement of localization in $U$. (This operator is denoted $E(U)$ in Wightman's treatment \cite{Wig:localizability}.)
The probability $P$ for localization in $U$ is the expectation value of this operator. For a (mixed) 1-particle state $\sigma$ this is
\begin{equation}
  P=\tr(\hat{\chi}_U \sigma) .
\end{equation}
The corresponding vectorial position operator $\hat{x}_i$ for $i=1,2,3$ is obtained by quantizing the components of the position vector $x_i$,
\begin{equation}
  \hat{x}_i\defeq \int\xd^3 x\, x_i\, a^\dagger_{t,x} \po_0 a_{t,x} .
\end{equation}
On the 1-particle Hilbert space $\cH^1$ this is precisely the Newton-Wigner position operator \cite{Wig:localizability}. In the following, our focus will be on the localized states rather than on the position operator.


\section{Probability density of position detection}
\label{sec:positionprob}

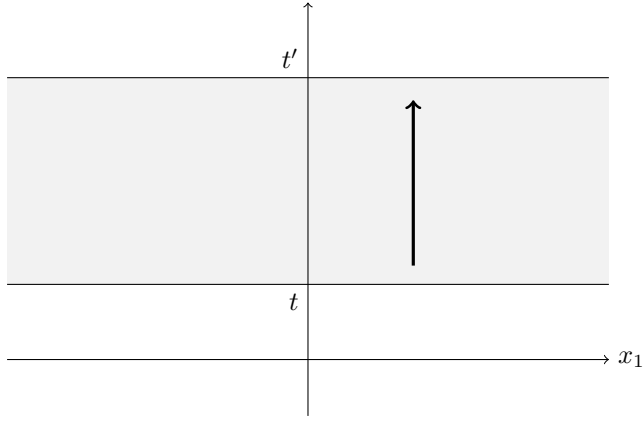
\begin{figure}
  \centering
  \begin{tikzpicture}
\fill[gray!10] (-4,0.75) rectangle (4,3.5);
\draw[black] (-4,0.75) -- (4,0.75);
\draw[black] (-4,3.5) -- (4,3.5);
\node at (0,3.5) [above left] {$t'$};
\node at (0,0.75) [below left] {$t$};
\draw[black,->] (-4,-0.25) -- (4,-0.25) node [right] {$x_1$};
\draw[black,->] (0,-1) -- (0,4.5); 
\draw[very thick,black,->] (1.4,1) -- (1.4,3.2);
\end{tikzpicture}
  \caption{Time evolution through the region between an initial spacelike hyperplane at time $t$ and a final spacelike hyperplane at time $t'$.}
  \label{fig:tevol}
\end{figure}

In the present section, we recall how the transition amplitude between Newton-Wigner states gives rise to a probability density for detection as a function of position. We consider evolution from an initial time $t$ to a final time $t'$, see Figure~\ref{fig:tevol}.
The transition amplitude $\rho$ between a one-particle state $\Psi_{t,x}$ at time $t$, localized at $x$ and a one-particle state $\Psi_{t',x'}$ at time $t'$, localized at $x'$ is given by the inner product \eqref{eq:rjip}, due to relation \eqref{eq:ccr},
\begin{equation}
    \rho(\Psi_{t,x} \tens \Psi_{t',x'})
    = \langle a^{\dagger}_{t',x'} \vac, a^{\dagger}_{t,x} \vac\rangle
    = \langle \vac, a_{t',x'} a^{\dagger}_{t,x} \vac\rangle
    = \langle \vac, [a_{t',x'}, a^{\dagger}_{t,x}] \vac\rangle
    = \{\Phi_{x}^t,\Phi_{x'}^{t'}\} .
    \label{eq:propampl}
\end{equation}
Note that the time-evolution is built into the creation and annihilation operators in our conventions and thus does not manifest in terms of a separate evolution operator. The corresponding probability density is thus,
\begin{equation}
  P(t',x'|t,x)=\left|\langle a^{\dagger}_{t',x'} \vac, a^{\dagger}_{t,x} \vac\rangle\right|^2=\left|\{\Phi_{x}^t,\Phi_{x'}^{t'}\}\right|^2 .
  \label{eq:propprob}
\end{equation}
In particular, $P(t',x'|t,x)$ has the interpretation of the probability density to find a particle at $x'$ that was initially localized at $x$.

From the definition \eqref{eq:locsol} and formula \eqref{eq:rjip} we obtain the integral expression
\begin{equation}
  \{\Phi_{x}^t,\Phi_{x'}^{t'}\}=\int\frac{\xd^3 k}{(2\pi)^3} e^{-\im(E(t'-t)-k(x'-x))} .
\end{equation}
We may notice that there is a simple relation to the Feynman propagator \eqref{eq:feynmanprop}.
It is then easy to see that the following equality holds (for $t'\ge t$)
\begin{equation}
\{\Phi_{x}^t,\Phi_{x'}^{t'}\} = - 2 \frac{\partial}{\partial (t'-t)} G_{\text{F}}(t,x;t',x').
\end{equation}
To evaluate this, it is convenient to introduce the invariant distance  $\sigma^2=(t'-t)^2-\|x'-x\|^2$ between $(t,x)$ and $(t',x')$. Then,
\begin{equation}
  \{\Phi_{x}^t,\Phi_{x'}^{t'}\}
  =\begin{dcases} -\frac{m^2 (t'-t)}{4\pi \sigma^2} H_2^{(2)}(m\sqrt{\sigma^2})
    & \text{if}\quad \sigma^2>0\quad\text{(timelike)} \\
    \frac{\im m^2 (t'-t)}{2\pi^2 \sigma^2} K_2(m\sqrt{-\sigma^2})
    & \text{if}\quad \sigma^2<0\quad\text{(spacelike)} .
  \end{dcases}
  \label{eq:propampleval}
\end{equation}
In the massless case this reduces to the simpler expression,
\begin{equation}
  \{\Phi_{x}^t,\Phi_{x'}^{t'}\}
  =-\frac{\im (t'-t)}{\pi^2 \sigma^4 } .
\end{equation}

\begin{figure}
  \centering
  \includegraphics[width=0.5\textwidth]{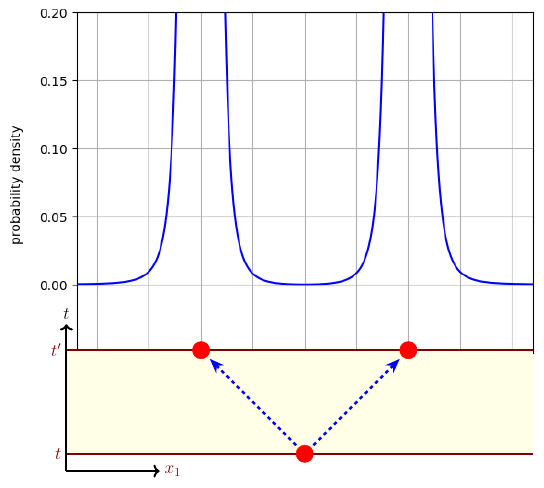}
  \caption{Spacetime representation of the localization problem (below) and of the associated probability density of detection (above). Here, $m=0$, $\Delta t=1$, $\Delta x=(\Delta x_1,0,0)$.}
  \label{fig:posdetection}
\end{figure}

A representation of the space-time setup and the probability density in the massless case is provided in Figure~\ref{fig:posdetection}. We set $t=0$, $x=(0,0,0)$ and $t'=1$. Supposing the particle is detected on the line $x_2=0,x_3=0$, the probability density as a function of $x_1$ is shown. Since the particle is massless, we expect it to travel at the speed of light and thus be detected either at $x_1=-1$ or $x_1=1$, depending on whether it moves to the left or right, see the arrows in the lower part of Figure~\ref{fig:posdetection}. This is where the invariant distance $\sigma$ vanishes and indeed singularities of the probability density appear, see the upper part of Figure~\ref{fig:posdetection}.

While the main feature of the probability distribution is its concentration on the light-cone, there is also a non-vanishing contribution outside the light-cone, including at $x_1'<-1$ and $x_1'>1$. For an initially perfectly localized classical particle at $x=(0,0,0)$ the latter would correspond to a regime of superluminal motion. We might ascribe this to an imperfection of the Newton-Wigner scheme, or perhaps more convincingly as arising from an intrinsic lack of perfect localizability in quantum field theory, see also comments in the Conclusions (Section~\ref{sec:conclusions}).

\begin{figure}
  \centering
  \includegraphics[width=0.5\textwidth]{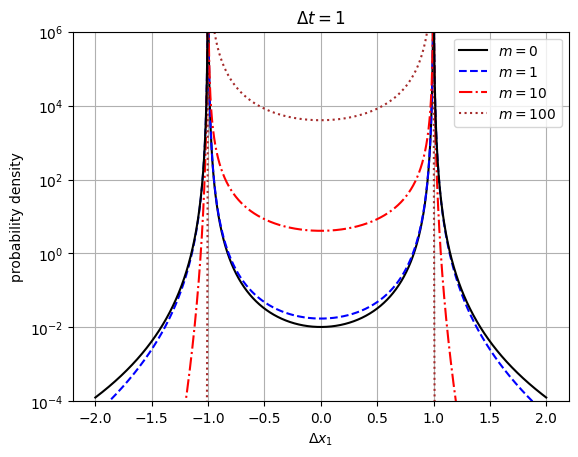}
  \caption{Probability density of detection as a function of distance for different values of the mass. Here, $\Delta t=1$, $\Delta x=(\Delta x_1,0,0)$.}
  \label{fig:pdenstint}
\end{figure}

We proceed to consider the massive case, see Figure~\ref{fig:pdenstint}. This shows the probability density for detection as a function of spatial displacement in one dimension (assuming no spatial displacement in the other directions). What we expect is that with increasing mass the expected effective velocity of the particle decreases more and more in comparison to the speed of light. This is exactly what we can read off from the graphs. As we have seen, the effective speed corresponds to the speed of light at $x_1'=-1$ and $x_1'=1$. For values in between the speed is lower, with rest at $x_1'=0$. The probability for encountering the particle at intermediate positions increases dramatically with an increase in the mass. However, a singular contribution on the light-cone remains a feature of the distributions for all values of the mass. At the same time, the probability to find a particle in the superluminal region at $x_1'<-1$ and $x_1'>1$ goes also down dramatically with increasing mass.


\section{Quantization on a timelike hyperplane}
\label{sec:Hstml}

In preparation for the construction of time-localized states we consider the quantization of massive Klein-Gordon theory on a timelike hyperplane. While we follow the procedure exactly as carried out in \cite{CoOe:evanescent}, to which we refer the reader for details, it is convenient for our present purposes to use a different parametrization of the classical phase space. What is more, we shall add a consideration of the energy flow through the hypersurface.

We consider a timelike hyperplane in Minkowski space characterized by a fixed value of the coordinate $x_1$, say at $x_1=z$. We recall that the phase space $L_z$, that is the space of germs of solutions of the equations of motion in a neighborhood of the hypersurface, splits into subspaces of \emph{propagating solutions} and of \emph{evanescent solutions}, $L_z=L_z^{\text{p}}\oplus L_z^{\text{e}}$.
We write $\tilde{x}=(x_2,x_3)$ and $\tilde{k}=(k_2,k_3)$ as a collective notation for positions and momenta in the two spatial coordinate directions tangential to the hypersurface.
The space $L_z^{\bC}$ of complexified solutions admits a parametrization as,
\begin{multline}
\phi(t,x_1,\tilde{x}) =  \int \frac{\xd^2\tilde{k}\, \xd E}{(2 \pi)^{3} 2k_1}\,\left(
    \left( \phi^{\Lr}_{E, \tilde{k}} d(E,k_1 x_1) + \phi^{\Ll}_{E, \tilde{k}}  \overline{d(E,k_1 x_1)} \right) e^{-\im E t + \im \tilde{k} \tilde{x}}
\right. \\
\left. + \left( \phi^{\Lcr}_{E, \tilde{k}} \overline{d(E,k_1 x_1)} + \phi^{\Lcl}_{E, \tilde{k}} d(E,k_1 x_1) \right) e^{\im E t - \im \tilde{k} \tilde{x}}
\right) .
\label{eq:kgtlp}
\end{multline}
The functions $d$ are combinations of exponentials as follows,
\begin{equation}
  d(E,w)\defeq \begin{cases} \exp(\im w) & \text{if}\; E>E_{\parallel} \\
    \cosh(w)+\im\sinh(w) & \text{if}\; E<E_{\parallel}
  \end{cases}
  \label{eq:dfunc}
\end{equation}
(Note that in \eqref{eq:kgtlp} $w=k_1 x_1$.) Here, $E_\parallel=\sqrt{\tilde{k}^2+m^2}$ and $k_1 = \sqrt{|E^2-\tilde{k}^2-m^2|}$. The integral in $\tilde{k}$ is over $\R^2$ and the integral in $E$ is over $\R^+$.
For $E>E_{\parallel}$ the solutions are \emph{propagating}. They are precisely the usual plane wave solutions as for the equal-time hypersurface (\ref{eq:kgmodes}), characterized by an oscillatory behavior in the $x_1$ direction, only slightly differently parametrized. For $E<E_{\parallel}$ the solutions are \emph{evanescent}. These solutions grow or decay exponentially in the direction perpendicular to the hypersurface.
Note that the parametrization~(\ref{eq:kgtlp}) is global in the sense of being independent of the location $z$ of the hyperplane. Consequently, the spaces $L_z$ for different $z$ are naturally identified. Mostly, we shall write $L$ instead of $L_z$ to emphasize this and also omit the index $z$ from other objects that can be globally defined in this sense. This is also in line with the standard practice in the time-evolution setting.

In the propagating sector, the coefficient functions $\phi^{\Lr}$ and $\phi^{\Lcr}$ correspond to right-moving plane wave solutions, i.e., waves that move in the positive $x_1$-direction as time increases. Correspondingly, the coefficient functions $\phi^{\Ll}$ and $\phi^{\Lcl}$ correspond to left-moving plane wave solutions. In the evanescent sector no corresponding behavior is apparent from the shape of the solutions. However, considering the energy flow through a timelike hyperplane reveals a notion of left- or right-moving solutions also in the evanescent sector. Thus, we consider the $T_{0 1}$-component of the energy-momentum tensor, integrated over the hypersurface. This yields the flux $F$ through the hypersurface to the left. It is conserved, i.e., independent of the $x_1$-coordinate of the hypersurface.
\begin{multline}
  F(\phi)=\int \xd t\,\xd^2 \tilde{x}\; T_{0 1}(t,x_1,\tilde{x})
  =\int \xd t\,\xd^2 \tilde{x}\; \partial_0 \phi(t,x_1,\tilde{x})\,
   \partial_1 \phi(t,x_1,\tilde{x}) \\
  =\int
  \frac{\xd^2\tilde{k}\,\xd E}{(2\pi)^3 2k_1} E
  \left(\phi^{\Ll}_{E,\tilde{k}}\phi^{\Lcl}_{E,\tilde{k}}
   -\phi^{\Lr}_{E,\tilde{k}}\phi^{\Lcr}_{E,\tilde{k}}\right) .
   \label{eq:eflux}
\end{multline}
This formula is valid both in the propagating and evanescent sectors. In the propagating sector, this just confirms the properties of being left- or right-moving as read off from the waveforms. In the evanescent sector, the parametrization \eqref{eq:kgtlp} and \eqref{eq:dfunc} is chosen so that precisely the same formula holds. That is, we may identify the coefficient functions $\phi^{\Lr}$ and $\phi^{\Lcr}$ as corresponding to right-moving solutions and $\phi^{\Ll}$ and $\phi^{\Lcl}$ as corresponding to left-moving solutions. However, this identification is to be taken with care as the components corresponding to the different coefficient functions are not translation invariant. This is in contrast to the propagating sector, where they are translation invariant. Given that the flux \eqref{eq:eflux} is conserved, i.e., translation invariant also in the evanescent sector, this might appear surprising. In order to deal with this mixing of left- and right-moving components under translation in the evanescent sector, we fix the notion of left- and right-moving to be in accordance with our parametrization \eqref{eq:kgtlp} at $x_1=0$. That is, we say that an evanescent solution is right-moving on a hypersurface at $x_1=z$, if translating the solution by $-z$ in the $x_1$-direction results in vanishing coefficients $\phi^{\Ll}$ and $\phi^{\Lcl}$, etc.

The symplectic form $\omega:L^{\C} \times L^{\C} \to \C$ is given by
\begin{multline}
  \omega(\phi,\eta)
  =\frac12 \int \xd t\,\xd^2 \tilde{x}
   \left(\phi(t,x_1,\tilde{x})\,
   \partial_1 \eta(t,x_1,\tilde{x})
   -\eta(t,x_1,\tilde{x})\,
   \partial_1 \phi(t,x_1,\tilde{x})\right) \\
  = \frac{\im}{2}\int
  \frac{\xd^2\tilde{k}\,\xd E}{(2\pi)^3 2k_1}
  \left(\phi^{\Lcr}_{E,\tilde{k}}\eta^{\Lr}_{E,\tilde{k}}
   +\phi^{\Ll}_{E,\tilde{k}}\eta^{\Lcl}_{E,\tilde{k}}
  -\phi^{\Lr}_{E,\tilde{k}}\eta^{\Lcr}_{E,\tilde{k}}
  -\phi^{\Lcl}_{E,\tilde{k}}\eta^{\Ll}_{E,\tilde{k}}\right) .
\end{multline}
As has been shown some time ago \cite{Oe:timelike}, quantization in the propagating sector can be performed quite analogous to quantization on a spacelike hypersurface. In particular, the choice of vacuum may be expressed in terms of a complex structure $J^{\Lp}:L^{\Lp,\C}\to L^{\Lp,\C}$. Here this is,
\begin{align}
  (J^{\Lp}(\phi))^{\Lr}_{E,\tilde{k}}
  =-\im \phi^{\Lr}_{E,\tilde{k}} \qquad
  (J^{\Lp}(\phi))^{\Ll}_{E,\tilde{k}}
  =\im \phi^{\Ll}_{E,\tilde{k}} , \nonumber \\
  (J^{\Lp}(\phi))^{\Lcr}_{E,\tilde{k}}
  =\im \phi^{\Lcr}_{E,\tilde{k}} \qquad
  (J^{\Lp}(\phi))^{\Lcl}_{E,\tilde{k}}
  =-\im \phi^{\Lcl}_{E,\tilde{k}} .
\end{align}
This yields the inner product,
\begin{equation}
  \{\phi,\eta\}^{\text{p}}
  =2\omega(\phi,J^{\Lp} \eta)+2\im\omega(\phi,\eta)
  =2\int_{E>E_\parallel}
  \frac{\xd^2\tilde{k}\,\xd E}{(2\pi)^3 2k_1}
  \left(\phi^{\Lr}_{E,\tilde{k}}\eta^{\Lcr}_{E,\tilde{k}}
  +\phi^{\Lcl}_{E,\tilde{k}}\eta^{\Ll}_{E,\tilde{k}}\right) ,
\end{equation}
determining the commutation relations \eqref{eq:ccr} between creation and annihilation operators in the propagating sector.

In the evanescent sector the standard quantization prescription fails as the Lagrangian subspaces corresponding to the physical vacuum are determined by a decay condition and thus real \cite{CoOe:vaclag}. Concretely, the subspaces to the left and right are respectively given by,
\begin{align}
  L^{\text{e},+} & = \{\phi\in L^{\text{e},\bC}:\;
   \phi^{\Lr}_{E,\tilde{k}}=-\im\phi^{\Ll}_{E,\tilde{k}},\;
   \phi^{\Lcr}_{E,\tilde{k}}=\im\phi^{\Lcl}_{E,\tilde{k}}\;
   \forall E,\tilde{k}\} ,
  \\
  L^{\text{e},-} & = \{\phi\in L^{\text{e},\bC}:\;
  \phi^{\Lr}_{E,\tilde{k}}=\im\phi^{\Ll}_{E,\tilde{k}},\;
  \phi^{\Lcr}_{E,\tilde{k}}=-\im\phi^{\Lcl}_{E,\tilde{k}}\;
  \forall E,\tilde{k}\} .
\end{align}
Instead, a novel twisted quantization prescription is required that was introduced in \cite{CoOe:locgenvac}. In the following, we only review the most relevant results of this quantization from \cite{CoOe:evanescent}, adapted to our present parametrization. Recall in particular, that in addition to a choice of vacuum a \emph{compatible real structure} on the complexified phase space $L^{\Le,\C}$ of the evanescent sector is required. This real structure $\alpha_u^{\Le}:L^{\Le,\C}\to L^{\Le,\C}$ is induced by a reflection map that depends on the position $x_1=u$ of a parallel reflection hypersurface. Here,
\begin{align}
  (\alpha^{\Le}_u(\phi))^{\Lr}_{E,\tilde{k}}
    =-\im\cosh(2 k_1 u)\overline{\phi^{\Lcl}_{E,\tilde{k}}} + \sinh(2 k_1 u)\overline{\phi^{\Lcr}_{E,\tilde{k}}}, \nonumber\\
  (\alpha^{\Le}_u(\phi))^{\Ll}_{E,\tilde{k}}
  =-\im\cosh(2 k_1 u)\overline{\phi^{\Lcr}_{E,\tilde{k}}} - \sinh(2 k_1 u)\overline{\phi^{\Lcl}_{E,\tilde{k}}} , \nonumber\\
  (\alpha^{\Le}_u(\phi))^{\Lcr}_{E,\tilde{k}}
   =-\im\cosh(2 k_1 u)\overline{\phi^{\Ll}_{E,\tilde{k}}} - \sinh(2 k_1 u)\overline{\phi^{\Lr}_{E,\tilde{k}}}, \nonumber\\
  (\alpha^{\Le}_u(\phi))^{\Lcl}_{E,\tilde{k}}
  =-\im\cosh(2 k_1 u)\overline{\phi^{\Lr}_{E,\tilde{k}}} + \sinh(2 k_1 u)\overline{\phi^{\Ll}_{E,\tilde{k}}} ,
\end{align}
The resulting inner product is given by,
\begin{equation}
  \{\phi,\eta\}^{\text{e}}  = \int_{E< E_\parallel}
  \frac{\xd^2\tilde{k}\,\xd E}{(2\pi)^3 2k_1}
  \left((\phi^{\Lr}_{E,\tilde{k}} +\im \phi^{\Ll}_{E,\tilde{k}})
    (\eta^{\Lcr}_{E,\tilde{k}} +\im \eta^{\Lcl}_{E,\tilde{k}})
    -(\phi^{\Lcr}_{E,\tilde{k}} -\im \phi^{\Lcl}_{E,\tilde{k}})
    (\eta^{\Lr}_{E,\tilde{k}} -\im \eta^{\Ll}_{E,\tilde{k}})
  \right) .
  \label{eq:rjipe}
\end{equation}
Crucially, creation and annihilation operators (as well as coherent states) are labeled not by elements of the real phase space $L^{\Le}$, but by element of $L^{\Le,u}\subseteq L^{\Le,\C}$. This is the subspace of the complexified phase space invariant under the real structure $\alpha^{\Le}_u$,
\begin{equation}
  L^{\Le,u}=\left\{\phi\in L^{\Le,\C}:\phi^{\Lr}_{E,\tilde{k}}+\im\phi^{\Ll}_{E,\tilde{k}}=e^{2 k_1 u} \left(\overline{\phi^{\Lcr}_{E,\tilde{k}}}-\im\overline{\phi^{\Lcl}_{E,\tilde{k}}}\right),
  \phi^{\Lr}_{E,\tilde{k}}-\im\phi^{\Ll}_{E,\tilde{k}}=-e^{-2 k_1 u} \left(\overline{\phi^{\Lcr}_{E,\tilde{k}}}+\im\overline{\phi^{\Lcl}_{E,\tilde{k}}}\right)  \right\} .
\end{equation}
In particular the inner product \eqref{eq:rjipe} determines the commutation relations \eqref{eq:ccr} of creation and annihilation operators in the evanescent sector, but with $\phi,\eta\in L^{\Le,u}$.

Since we need to construct particle states that are in correspondence to elements of the classical phase space, we use a canonical identification map $I^{\Le,u}:L^{\Le,\C}\to L^{\Le,\C}$ that maps the real phase space $L^{\Le}$ to $L^{\Le,u}$. This map takes the form,
\begin{align}
  (I^{\Le,u}(\phi))^{\Lr}_{E,\tilde{k}}
  & = \frac{1}{\sqrt{2}} \left((1+\sinh(2 k_1 u))
    \phi^{\Lr}_{E,\tilde{k}} -\im\cosh(2 k_1 u)\phi^{\Ll}_{E,\tilde{k}}\right), \nonumber\\
  (I^{\Le,u}(\phi))^{\Ll}_{E,\tilde{k}}
  & = \frac{1}{\sqrt{2}} \left((1-\sinh(2 k_1 u))
    \phi^{\Ll}_{E,\tilde{k}} -\im\cosh(2 k_1 u)\phi^{\Lr}_{E,\tilde{k}}\right), \nonumber\\
  (I^{\Le,u}(\phi))^{\Lcr}_{E,\tilde{k}}
  & = \frac{1}{\sqrt{2}} \left((1-\sinh(2 k_1 u))
    \phi^{\Lcr}_{E,\tilde{k}} -\im\cosh(2 k_1 u)\phi^{\Lcl}_{E,\tilde{k}}\right), \nonumber\\
  (I^{\Le,u}(\phi))^{\Lcl}_{E,\tilde{k}}
  & = \frac{1}{\sqrt{2}} \left((1+\sinh(2 k_1 u))
    \phi^{\Lcl}_{E,\tilde{k}} -\im\cosh(2 k_1 u)\phi^{\Lcr}_{E,\tilde{k}}\right) . \label{eq:eproj}
\end{align}


\section{Time-localized states and the Time Operator}
\label{sec:timeop}

We proceed to perform a construction quite analogous to the Newton-Wigner states and corresponding operator, but with localization on the timelike hyperplane at $x_1=z$.
In the propagating sector consider the solution $\Phi_{t,\tilde{y}}\in L_{z}^{\Lp}$ with $t\in\R$ a time and $\tilde{y}\in\R^2$ a position on the plane spanned by $x_2$ and $x_3$, given as follows.
\begin{align}
  (\Phi_{t,\tilde{y}}^z)^{\Lr}_{E,\tilde{k}}  & =\sqrt{k_1} e^{\im(E t- \tilde{k} \tilde{y})} e^{-\im k_1 z}, \quad\text{for}\; E>E_\parallel, \label{eq:psolr}\\
  (\Phi_{t,\tilde{y}}^z)^{\Ll}_{E,\tilde{k}}
  & =\sqrt{k_1} e^{\im(E t- \tilde{k} \tilde{y})}  e^{\im k_1 z}, \quad\text{for}\; E>E_\parallel, \label{eq:psoll} \\
  (\Phi_{t,\tilde{y}}^z)^{\Lcr}_{E,\tilde{k}}  & =\sqrt{k_1} e^{-\im(E t- \tilde{k} \tilde{y})} e^{\im k_1 z}, \quad\text{for}\; E>E_\parallel, \label{eq:psolcr} \\
  (\Phi_{t,\tilde{y}}^z)^{\Lcl}_{E,\tilde{k}}
  & =\sqrt{k_1} e^{-\im(E t- \tilde{k} \tilde{y})} e^{-\im k_1 z}, \quad\text{for}\; E>E_\parallel . \label{eq:psolcl}
\end{align}
In the evanescent sector, the relevant solution must live in the space $L_z^{\Le,u}$, which depends on the location $x_1=u$ of the reflection hyperplane that determines the $\alpha$-Kähler quantization performed in Section~\ref{sec:Hstml}. The most natural way to fix this quantization ambiguity is to set $u=z$ for each hyperplane. That is, we always take the reflection hyperplane to be same hyperplane as that of the respective space of solutions and quantum state space.
To obtain solutions in  $L_z^{\Le,z}$, we start with real solutions in $L_z^{\Le}$ and project them via the map $I^{\Le,z}$, of equation \eqref{eq:eproj}. We do this at $z=u=0$, where the required form of the coefficients is obvious and simply coincides formally with those of the propagating sector, expressions \eqref{eq:psolr}--\eqref{eq:psolcl}. Then, we translate the solutions from $0$ to the desired value $z$ of the $x_1$-coordinate.
\begin{align}
  (\Phi_{t,\tilde{y}}^z)^{\Lr}_{E,\tilde{k}}  &
   =e^{-\im\pi/4}\sqrt{k_1} e^{\im(E t- \tilde{k} \tilde{y})} d(k_1 z), \quad\text{for}\; E<E_\parallel, \\
  (\Phi_{t,\tilde{y}}^z)^{\Ll}_{E,\tilde{k}}  &
   =e^{-\im\pi/4}\sqrt{k_1} e^{\im(E t- \tilde{k} \tilde{y})} \overline{d(k_1 z)}, \quad\text{for}\; E<E_\parallel, \\
  (\Phi_{t,\tilde{y}}^z)^{\Lcr}_{E,\tilde{k}}  &
   =e^{-\im\pi/4}\sqrt{k_1} e^{-\im(E t- \tilde{k} \tilde{y})} \overline{d(k_1 z)}, \quad\text{for}\; E<E_\parallel, \\
  (\Phi_{t,\tilde{y}}^z)^{\Lcl}_{E,\tilde{k}}  &
   =e^{-\im\pi/4}\sqrt{k_1} e^{-\im(E t- \tilde{k} \tilde{y})} d(k_1 z), \quad\text{for}\; E<E_\parallel .
\end{align}

We have the following orthogonality and completeness relations,
\begin{gather}
  \{\Phi_{t',\tilde{y}'}^z,\Phi_{t,\tilde{y}}^z\}=\delta(t-t')\delta^2(\tilde{y}-\tilde{y}') , \label{eq:tdeltanorm} \\
  \{\xi,\eta\}=\int\xd t\,\xd^2\tilde{y}\, \{\xi,\Phi_{t,\tilde{y}}^z\} \{\Phi_{t,\tilde{y}}^z,\eta\}
  \qquad\forall\,\xi,\eta\in L_z^z .
  \label{eq:tcompl}
\end{gather}
Crucially, restricting to either propagating or evanescent modes only on the right-hand side introduces a projection to the corresponding subspace of $L_z^z$. In particular, this would correspond to a cut-off in the energy-momentum space of the solutions on the hyperplane and thus break completeness in particular for time-localized solutions, see also Section~\ref{sec:seppe}.

Upon $\alpha$-Kähler quantization, the completeness relation yields a corresponding completeness relation on the 1-particle Hilbert subspace $\cH_z^{u,1}\subseteq \cH_z^{u}$ of states at $x_1=z$. We can express this in terms of creation and annihilation operators as follows,
\begin{equation}
  \id_{z}^1=\int\xd t\,\xd^2\tilde{y}\, a^\dagger_{z,t,\tilde{y}} \po_0 a_{z,t,\tilde{y}} .
\end{equation}
We use the notation $a_{z,t,\tilde{y}}\defeq a_{\Phi_{t,\tilde{y}}^z}$. Again, $\po_0$ denotes the projector onto the vacuum state. With this we have the quantization map from real valued functions on $\R\times\R^2$ (time times space in the plane) to self-adjoint operators on $\cH_z^u$,
\begin{equation}
  \hat{f}\defeq \int\xd t\,\xd^2\tilde{y}\, f(t,\tilde{y})\, a^\dagger_{z,t,\tilde{y}} \po_0 a_{z,t,\tilde{y}} .
\end{equation}
For non-negative functions $f$, $\hat{f}$ is positive, so this defines a POVM on $\cH_z^{u,1}$.
Analogous to the Newton-Wigner setting, we have the following probability interpretation. For a Lebesgue measurable subset $U\subseteq \R\times \R^2$ the probability of measuring the particle in $U$, given the 1-particle state $\sigma$ on the left-hand side is,
\begin{equation}
  P=\tr_z(\hat{\chi}_U \sigma),
\end{equation}
if $\tr_z(\sigma)=1$. We are assuming here that we have no knowledge at all about the state of affairs on the right-hand side of the hyperplane.

Finally, we can define the corresponding vectorial position operator $\hat{\tilde{y}}_i$ with $i=2,3$ and the time operator $\hat{t}$. The latter, being of principal interest here, is given by,
\begin{equation}
  \hat{t}\defeq \int\xd t\,\xd^2\tilde{y}\, t\, a^\dagger_{z,t,\tilde{y}} \po_0 a_{z,t,\tilde{y}} .
\end{equation}
The corresponding expectation value of the operator $\hat{t}$ indeed has the interpretation of the expected incidence time of a particle on the screen $x_1=z$. Again, this is assuming a state on the left-hand side and no knowledge about the right-hand side. In the following, our focus will be on the time-localized states rather than on the time operator.


\section{The time-of-arrival problem}
\label{sec:toa}

\begin{figure}
  \centering
  \begin{tikzpicture}
\filldraw[gray!10] (1,-1) rectangle (3,3);
\draw[->] (-1,0) -- (5,0) node [right] {$x_1$};
\draw[->] (0,-1) -- (0,3) node [left] {$t$};
\draw[-] (1,-1) -- (1,3);
\draw[-] (3,-1) -- (3,3);
\node at (1,0) [below left] {$z$};
\node at (3,0) [below right] {$z'$};
\draw[very thick,->] (1.2,1.5) -- (2.8,1.5);
\end{tikzpicture}
  \caption{Spatial evolution through the region between an initial (left) timelike hyperplane at $x_1=z$ and a final (right) timelike hyperplane at $x_1=z'$.}
  \label{fig:spevol}
\end{figure}
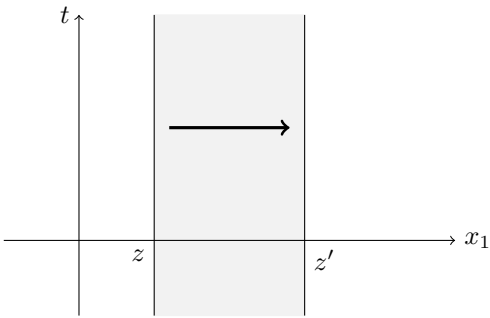

Our first application of the time-localized states just introduced is the time-of-arrival problem. We formulate this as follows: 
We wish to predict the probability distribution of the arrival time and location on the timelike hyperplane at $x_1=z'$ for a particle emitted on the timelike hyperplane at $x_1=z$ at a specific time and location. We can think of this as a \emph{spatial evolution} problem, see Figure~\ref{fig:spevol}. The figure looks just like the setup for the propagation of a particle between spacelike hyperplanes, see Figure~\ref{fig:tevol}, except rotated by 90° and thereby swapping one spatial direction with the temporal direction. Making this intuition of a rotated picture rigorous is precisely the basic idea of the present approach.

\subsection{Bidirectional incidence problem}
\label{sec:bincprob}

The transition amplitude $\rho$ between a one-particle state $\Psi_{z,t,\tilde{y}}$ at $x_1=z$, at time $t$ and location $\tilde{y}$ and a one-particle state $\Psi_{z',t',\tilde{y}'}$ at $x_1=z'$, at time $t'$ and location $\tilde{y}'$ is given by the inner product \eqref{eq:rjip}, as follows,
\begin{equation}
    \rho(\Psi_{z,t,\tilde{y}}\tens\Psi_{z',t',\tilde{y}'})
    =\{\Phi_{t,\tilde{y}}^z,\Phi_{t',\tilde{y}'}^{z'}\} .
    \label{eq:propamplt}
\end{equation}
This appears to be completely analogous to the amplitude \eqref{eq:propampl} for the propagation between spacelike hyperplanes. However, in the present case the arguments of the inner product live in different subspaces of the space of complexified germs of solutions $L^\C$. This is due to the different choices of reflection hypersurfaces determining the quantizations on the timelike hyperplanes at $z$ and $z'$ respectively. In particular, we have $\Phi_{t,\tilde{y}}^z\in L^{\Lp}\oplus L^{\Le,z}$ and $\Phi_{t',\tilde{y}'}^{z'}\in L^{\Lp}\oplus L^{\Le,z'}$. The derivation of equation \eqref{eq:propamplt} is thus more involved than that of \eqref{eq:propampl}, but follows from the results of the work \cite{CoOe:evanescent}.\footnote{More precisely, equation (59) of \cite{CoOe:evanescent} provides an analogous equation for coherent states. Adapted to the present notation, this is $\rho_{[z,z']}(\coh_{\xi}\tens\coh_{\xi'})=\exp\left(\frac12 \{\xi,\xi'\}\right)$. Taking functional derivatives using the relation $\Psi_{\xi}=\sqrt{2}\frac{\partial}{\partial \lambda}\coh_{\lambda\xi} \big|_{\lambda=0}$ yields a corresponding result for one-particle states. Finally, setting $\xi=\Phi_{t,\tilde{y}}^z$ and $\xi'=\Phi_{t',\tilde{y}'}^{z'}$ yields the stated result.}

The probability density for the incidence of a particle at time $s'$ and location $\tilde{y}'$ on the timelike hyperplane at $x_1=z'$ given a particle incidence at time $s$ and location $\tilde{y}$ on the timelike hyperplane at $x_1=z$ is
\begin{equation}
    P(z',t',\tilde{y}'|z,t,\tilde{y})=\left| \{\Phi_{t,\tilde{y}}^z,\Phi_{t',\tilde{y}'}^{z'}\}\right|^2 .
    \label{eq:propprobt}
\end{equation}
Again, this appears to be completely analogous to the Newton-Wigner setting, see expression \eqref{eq:propprob}. However, the probability interpretation in the present context cannot be derived within the standard formulation of quantum theory. Instead, it relies on the more general framework of the positive formalism \cite{Oe:posfound}. For the present context, this is explained in Section~5 of the work \cite{CoOe:evanescent}. For the purpose of this paper, where we are only interested in the functional dependence of an un-normalized probability density, expression \eqref{eq:propprobt} is completely adequate.

Again, the amplitude is related in a simple way to the Feynman propagator \eqref{eq:feynmanprop}. In this case the relation is,
\begin{equation}
\{\Phi_{t,\tilde{y}}^z,\Phi_{t',\tilde{y}'}^{z'}\} = 2 \frac{\partial}{\partial (z'-z)} G_{\text{F}}(t,\tilde{y},z;t',\tilde{y}',z').
\end{equation}
Introducing the invariant distance $\sigma^2=(t'-t)^2-\|\tilde{y}'-\tilde{y}\|^2-(z'-z)^2$, evaluation of the amplitude yields (see equations \eqref{eq:ampl-tm} and \eqref{eq:ampl-sp} of the appendix),
\begin{equation}
    \{\Phi_{t,\tilde{y}}^z,\Phi_{t',\tilde{y}'}^{z'}\}
    =\begin{dcases} \frac{m^2 (z'-z)}{4\pi \sigma^2} H_2^{(2)}(m\sqrt{\sigma^2})
    & \text{if}\quad \sigma^2>0\quad\text{(timelike)} \\
    -\frac{\im m^2 (z'-z)}{2\pi^2 \sigma^2} K_2(m\sqrt{-\sigma^2})
    & \text{if}\quad \sigma^2<0\quad\text{(spacelike)} .
  \end{dcases}
  \label{eq:biampl}
\end{equation}
This is identical to the corresponding expression \eqref{eq:propampleval} for the amplitude between Newton-Wigner states, except for the replacement of the time difference $t'-t$ by the spatial distance $z'-z$. This is also true for the massless expression (see equation \eqref{eq:ampl-massless} of the appendix)
\begin{equation}
    \{\Phi_{t,\tilde{y}}^z,\Phi_{t',\tilde{y}'}^{z'}\}
    =\frac{\im (z'-z)}{\pi^2 \sigma^4 } .
    \label{eq:biampl0}
\end{equation}

\begin{figure}
  \centering
  \includegraphics[height=0.5\textwidth]{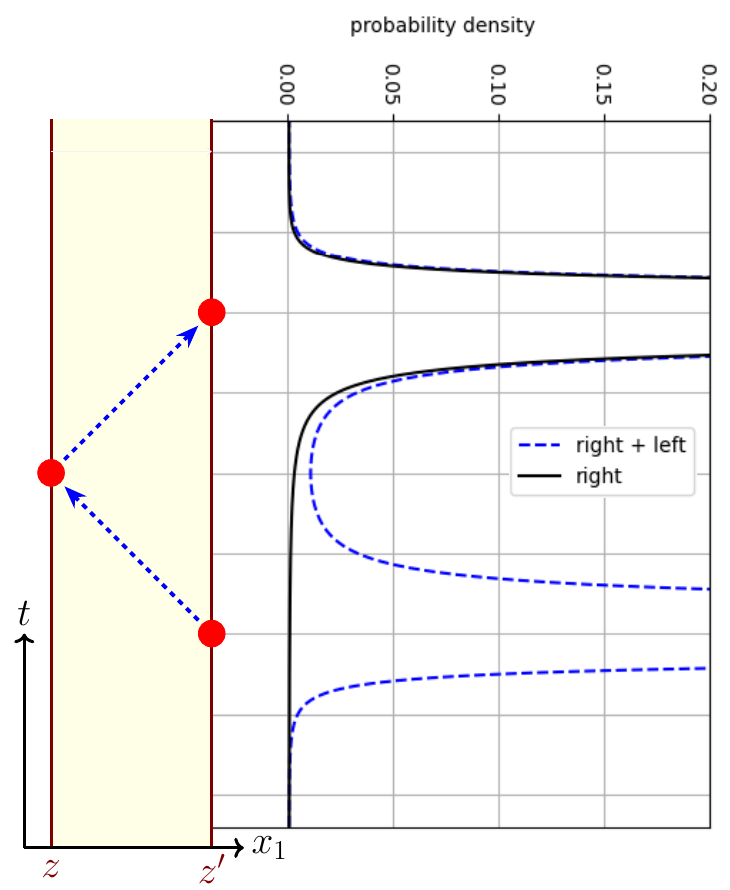}
  \caption{Graphical representation of the temporal incidence problem (left) and of the associated probability density of detection (right), blue dashed curve. The black curve corresponds to the arrival problem. Here, $m=0$, $\Delta z=1$, $\tilde{y}=\tilde{y}'=(0,0)$.}
  \label{fig:timedetection}
\end{figure}

A representation of the spacetime setup and the probability density in the massless case is provided in Figure~\ref{fig:timedetection}. For comparability with the corresponding Figure~\ref{fig:posdetection} of the Newton-Wigner setup, the orientations of space and time axes are the same. As expected, at first sight, the two spacetime setups look like 90° rotated copies of one another. What is more, the probability densities also look alike with two peaks corresponding to two singularities (compare the blue dashed line). However, in the present setting, in contrast to Newton-Wigner, the two singularities have quite different meanings. Indeed, for the time-of-arrival problem, we only expect one singularity. Namely, when a particle emitted at $z=0$, $\tilde{y}=(0,0)$ and time $t=0$ hits the screen at $z'=1$ and time $t'=1$, moving with the speed of light, (we set $\tilde{y}'=(0,0)$). This is the trajectory represented by the upper right-pointing arrow in Figure~\ref{fig:timedetection} and the upper peak of the probability density (blue dashed line). However, in our construction of time-localized states, we do not impose any restriction on the direction of motion of the particle perpendicular to the hypersurface. This is in stark contrast to particle states on a spacelike hypersurface, which always "move to the future". Consequently, we cannot a priori qualify the event on the left hypersurface as an "emission". It might as well be an "absorption". Indeed, in this latter case we expect a particle to emanate at an earlier time from the hypersurface on the right-hand side. More precisely, it should be emitted from there at $t'=-1$ in order to travel at the speed of light. This is precisely what the lower peak in Figure~\ref{fig:timedetection} represents, see also the lower left-pointing arrow representing the particle trajectory. Since we are not distinguishing between arrival and departure, we call the present setup, the \emph{incidence problem}.

As in the Newton-Wigner setting, the probability distribution does not vanish away from the light-cone. In particular, it has a non-vanishing contribution also in the superluminal regime that corresponds here to the intermediate times $-1< t' < 1$. Again, we attribute this to intrinsic quantum uncertainty.

\subsection{Unidirectional arrival problem}
\label{sec:unidir}

We are now ready to consider the original time-of-arrival problem. To this end we modify the scenario just considered by imposing the additional condition that the particle originates on the left-hand hypersurface at $z$ and is detected on the right-hand hypersurface at $z'$, excluding the opposite possibility. The time-localized states previously introduced do not distinguish a direction of motion of the particle perpendicular to the hypersurface. We thus introduce an additional quantum number to encode this distinction. The previously defined states are then recovered as sums over the two possible values of this quantum number.

Due to our parametrization of solutions in these terms, the distinction between left- and right-moving components is straightforward in the propagating sector. Separating out the components of $\Phi_{t,\tilde{y}}^z$ and denoting them $\Phi_{\Ll,t,\tilde{y}}^z$ and $\Phi_{\Lr,t,\tilde{y}}^z$, we obtain straightforwardly from expressions \eqref{eq:psolr}--\eqref{eq:psolcl},
\begin{align}
    (\Phi_{\Lr,t,\tilde{y}}^z)^{\Lr}_{E,\tilde{k}}  & =\sqrt{k_1} e^{\im(E t- \tilde{k} \tilde{y})} e^{-\im k_1 z}, &
    (\Phi_{\Lr,t,\tilde{y}}^z)^{\Ll}_{E,\tilde{k}}  = 0,
    \quad\text{for}\; E>E_\parallel, 
    \label{eq:modeR}
    \\
    (\Phi_{\Lr,t,\tilde{y}}^z)^{\Lcr}_{E,\tilde{k}}  & =\sqrt{k_1} e^{-\im(E t- \tilde{k} \tilde{y})} e^{\im k_1 z}, &
    (\Phi_{\Lr,t,\tilde{y}}^z)^{\Lcl}_{E,\tilde{k}}  = 0,
    \quad\text{for}\; E>E_\parallel, 
    \label{eq:modeRbar}
    \\
    (\Phi_{\Ll,t,\tilde{y}}^z)^{\Ll}_{E,\tilde{k}}
    & =\sqrt{k_1} e^{\im(E t- \tilde{k} \tilde{y})}  e^{\im k_1 z}, &
    (\Phi_{\Ll,t,\tilde{y}}^z)^{\Lr}_{E,\tilde{k}}  = 0,
    \quad\text{for}\; E>E_\parallel, 
    \label{eq:modeL}
    \\
    (\Phi_{\Ll,t,\tilde{y}}^z)^{\Lcl}_{E,\tilde{k}}
    & =\sqrt{k_1} e^{-\im(E t- \tilde{k} \tilde{y})} e^{-\im k_1 z}, &
    (\Phi_{\Ll,t,\tilde{y}}^z)^{\Lcr}_{E,\tilde{k}}  = 0,
    \quad\text{for}\; E>E_\parallel .
    \label{eq:modeLbar}
\end{align}

In the evanescent sector we proceed in a manner analogous to the way we obtained the full localized solutions. We start with the same solution as before, in the real solution space $L_0^{\Le}$ at $u=0$. However, this time we restrict in this same space to either the right-moving or the left-moving components. Then we map via $I^{\Le,0}$ to $L_0^{\Le,0}$, and finally we translate by $z$ to the right to obtain a solution in $L_z^{\Le,z}$. This yields the right-moving solutions,
\begin{align}
    (\Phi_{\Lr,t,\tilde{y}}^z)^{\Lr}_{E,\tilde{k}}  &
     =\sqrt{k_1/2}\, e^{\im(E t- \tilde{k} \tilde{y})} e^{k_1 z}, \quad\text{for}\; E<E_\parallel, 
     \label{eq:modeRev}\\
    (\Phi_{\Lr,t,\tilde{y}}^z)^{\Ll}_{E,\tilde{k}}  &
     =-\im \sqrt{k_1/2}\, e^{\im(E t- \tilde{k} \tilde{y})} e^{k_1 z}, \quad\text{for}\; E<E_\parallel, 
     \label{eq:modeLev}\\
    (\Phi_{\Lr,t,\tilde{y}}^z)^{\Lcr}_{E,\tilde{k}}  &
     =\sqrt{k_1/2}\, e^{-\im(E t- \tilde{k} \tilde{y})} e^{-k_1 z}, \quad\text{for}\; E<E_\parallel, 
     \label{eq:modeRbarev}\\
    (\Phi_{\Lr,t,\tilde{y}}^z)^{\Lcl}_{E,\tilde{k}}  &
     =-\im \sqrt{k_1/2}\, e^{-\im(E t- \tilde{k} \tilde{y})} e^{-k_1 z}, \quad\text{for}\; E<E_\parallel .
     \label{eq:modeLbarev}
\end{align}
The left-moving solutions are,
\begin{align}
    (\Phi_{\Ll,t,\tilde{y}}^z)^{\Lr}_{E,\tilde{k}}  &
     =-\im \sqrt{k_1/2}\, e^{\im(E t- \tilde{k} \tilde{y})} e^{-k_1 z}, \quad\text{for}\; E<E_\parallel, \\
    (\Phi_{\Ll,t,\tilde{y}}^z)^{\Ll}_{E,\tilde{k}}  &
     =\sqrt{k_1/2}\, e^{\im(E t- \tilde{k} \tilde{y})} e^{-k_1 z}, \quad\text{for}\; E<E_\parallel, \\
    (\Phi_{\Ll,t,\tilde{y}}^z)^{\Lcr}_{E,\tilde{k}}  &
     =-\im \sqrt{k_1/2}\, e^{-\im(E t- \tilde{k} \tilde{y})} e^{k_1 z}, \quad\text{for}\; E<E_\parallel, \\
    (\Phi_{\Ll,t,\tilde{y}}^z)^{\Lcl}_{E,\tilde{k}}  &
     =\sqrt{k_1/2}\, e^{-\im(E t- \tilde{k} \tilde{y})} e^{k_1 z}, \quad\text{for}\; E<E_\parallel .
\end{align}
We may verify that left-moving and right-moving modes sum to the modes defined in Section~\ref{sec:timeop}, $\Phi_{t,\tilde{y}}^z=\Phi_{\Lr,t,\tilde{y}}^z+\Phi_{\Ll,t,\tilde{y}}^z$, as required.

We proceed to consider the inner product between localized and directed solutions at different locations with $z'>z$. As is to be expected, the inner product between a left-moving particle state on one hypersurface and a right-moving particle state on the other hypersurface vanishes. However, to obtain analytic expressions for the non-trivial amplitudes turns out to be difficult. We thus start with the massless case. Set $\Delta x$ to be the undirected spatial distance given by $(\Delta x)^2=(z'-z)^2+\|\tilde{y}'-\tilde{y}\|^2$ and $\Delta t=t'-t$ the directed temporal difference. This is (see equations \eqref{eq:rightampl} and \eqref{eq:leftampl} of the appendix)
\begin{align}
    \{\Phi_{\Lr,t,\tilde{y}}^z,\Phi_{\Lr,t',\tilde{y}'}^{z'}\} &
    = \frac{\im (z'-z)   \left( 2 \Delta x - \Delta t \right)  }{(2\pi)^2   \left(  \Delta x - \Delta t \right)^2  \left(\Delta x\right)^3 },
    \label{eq:amplr} \\
    \{\Phi_{\Ll,t,\tilde{y}}^z,\Phi_{\Ll,t',\tilde{y}'}^{z'}\} &
    = \frac{\im (z'-z)   \left( 2 \Delta x + \Delta t \right)  }{(2\pi)^2   \left(  \Delta x + \Delta t \right)^2  \left(\Delta x\right)^3 }.
    \label{eq:ampll}
\end{align}
The probability densities for the corresponding processes are again obtained by taking the modulus square of the amplitudes, compare relation~\eqref{eq:propprobt}.

For definiteness, we limit ourselves in the following to right-moving particles. The case of left-moving particles is completely analogous. Thus, consider a right-moving particle emitted at time $t=0$ and at $z=0$ and $\tilde{y}=(0,0)$. Again, we detect at $z'=1$ and fix $\tilde{y}'=(0,0)$. The probability density of detection as a function of the time $t'$ is then given by the black continuous curve in Figure~\ref{fig:timedetection}. As is easy to see, this curve coincides perfectly with the dashed blue curve for values of time near $t=1$ and above. This confirms that the singularity at $t=1$ is attributable exclusively to a particle emitted at $z=0$ and time $t=0$ arriving at $z'=1$ at time $t=1$. It corresponds to the spacetime trajectory indicated by only the upward right-pointing arrow. We are indeed considering the ordinary time-of-arrival problem.

In the following we consider the time-of-arrival problem for massive particles. The amplitudes are given by the following integral expressions,
\begin{align}
  \{\Phi_{\Lr,t,\tilde{y}}^z,\Phi_{\Lr,t',\tilde{y}'}^{z'}\}
  & = \int
  \frac{\xd^2\tilde{k}\,\xd E}{(2\pi)^3}\,
  e^{-\im (E(t'-t)-\tilde{k}(\tilde{y}'-\tilde{y}))} e^{\im \underline{k}_1 (z'-z)} 
  \label{eq:aR}\\
  \{\Phi_{\Ll,t,\tilde{y}}^z,\Phi_{\Ll,t',\tilde{y}'}^{z'}\}
  & = \int
  \frac{\xd^2\tilde{k}\,\xd E}{(2\pi)^3}\,
  e^{\im (E(t'-t)-\tilde{k}(\tilde{y}'-\tilde{y}))} e^{\im \underline{k}_1 (z'-z)} .
  \label{eq:aL}
\end{align}
In contrast to the conventions of Section~\ref{sec:Hstml} we let $\underline{k}_1$ here be determined by analytical continuation when transitioning from the propagating to the evanescent sector. That is,
\begin{equation}
  \underline{k}_1\defeq \begin{cases}
    \sqrt{E^2-\tilde{k}^2-m^2} & \text{if} \, E^2> \tilde{k}^2-m^2, \\
    \im\sqrt{\tilde{k}^2+m^2-E^2} & \text{if} \, E^2< \tilde{k}^2-m^2 .
  \end{cases}
\end{equation}
Unfortunately, these integrals are difficult to evaluate in the massive case $m>0$. However, we can infer from the massless case that the probability density for the arrival problem is essentially the same as that for the incidence problem once we restrict consideration to positive times $t>0$ away from $t=0$. In this regime we may use the expressions \eqref{eq:biampl} for the bidirectional amplitudes as very good approximations to the unidirectional amplitudes of the problem at hand.

\begin{figure}
  \centering
  \includegraphics[width=0.5\textwidth]{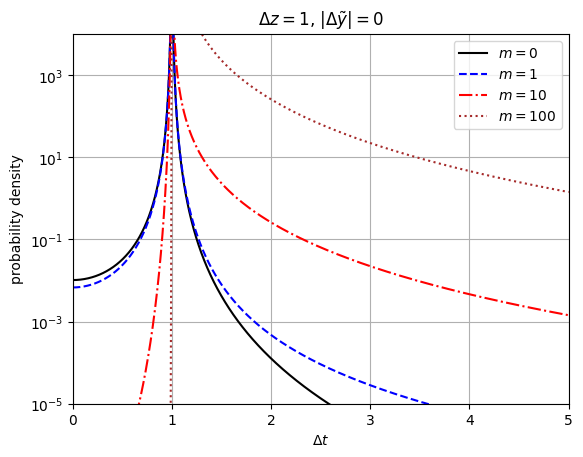}
  \caption{Probability density as a function of time $t'$ in the time-of-arrival problem for different masses. Here, $z=0$, $\tilde{y}=(0,0)$, $t=0$ and $z'=1$, $\tilde{y}'=(0,0)$.}
  \label{fig:pdens-arrival}
\end{figure}

Figure~\ref{fig:pdens-arrival} shows the probability density as a function of arrival time $\Delta t$ for different values of the mass $m$. Here, the distance is fixed to $\Delta x=1$, and we assume arrival without displacement in the $y$-directions. As can be clearly seen, the probability for later arrival increases markedly with increasing mass. That is, the expected effective velocity of the particle decreases with mass, as expected. At the same time, the probability of apparent superluminal propagation, i.e., at $\Delta t<1$ decreases sharply with increasing mass.
As commented before, the formulas used for generating the plots are those of incidence problem rather than those of the arrival problem. The error caused by this is only a very slight upwards displacement near $\Delta t=0$ for the $m=0$ and $m=1$ curves.

\begin{figure}
  \begin{tabular}{ccc}
    \includegraphics[width=0.3\textwidth]{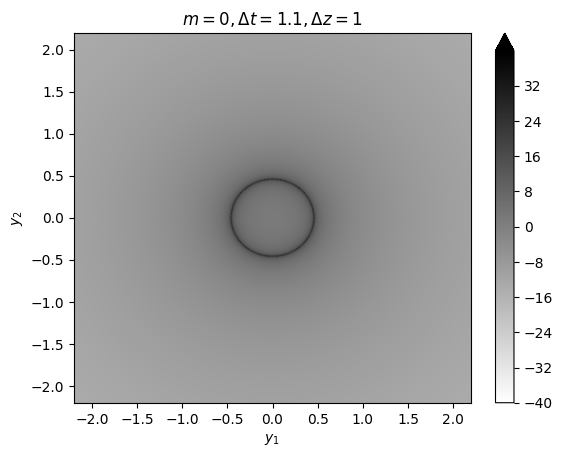} &
    \includegraphics[width=0.3\textwidth]{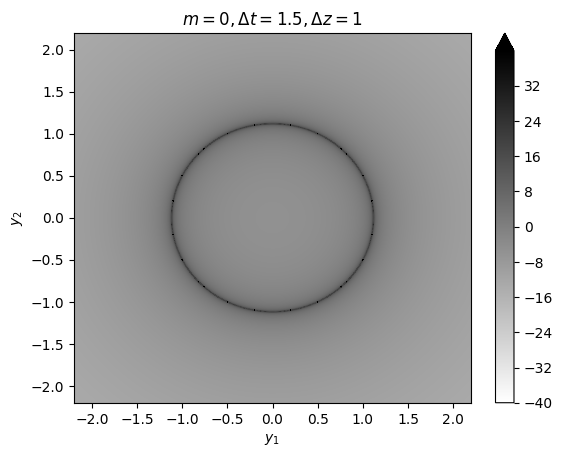} &
    \includegraphics[width=0.3\textwidth]{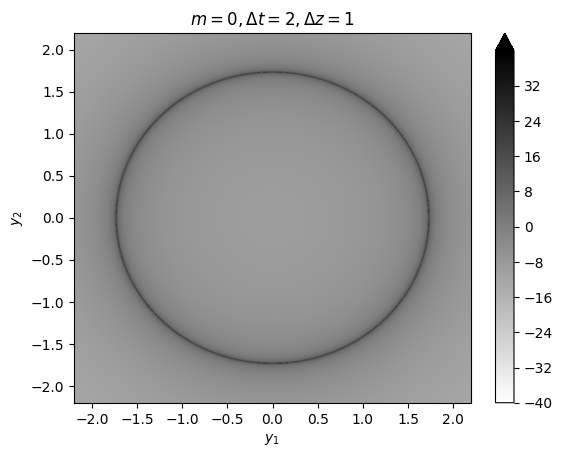} \\
    \includegraphics[width=0.3\textwidth]{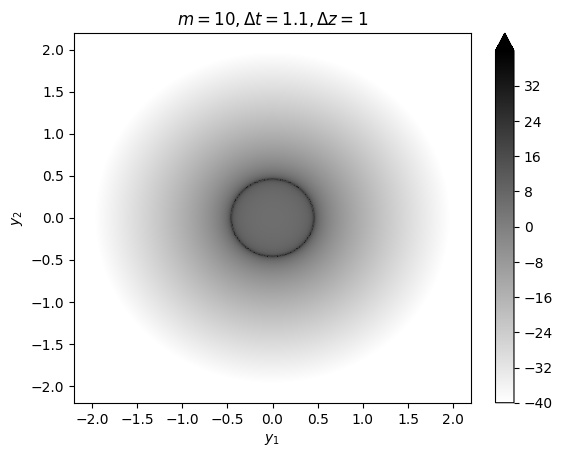} &
    \includegraphics[width=0.3\textwidth]{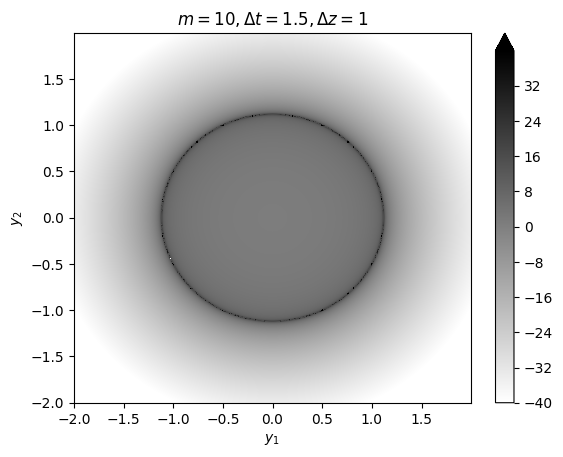} &
    \includegraphics[width=0.3\textwidth]{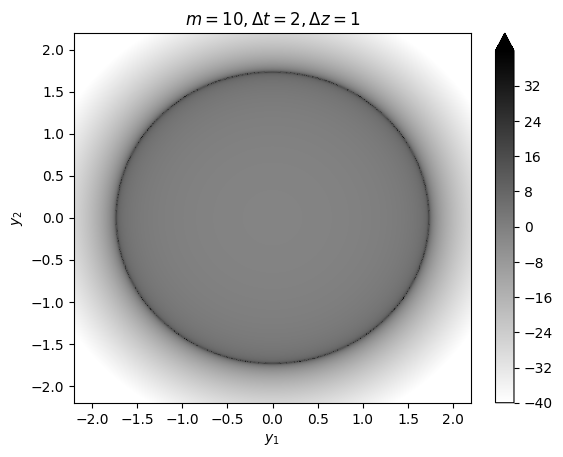} \\
    \includegraphics[width=0.3\textwidth]{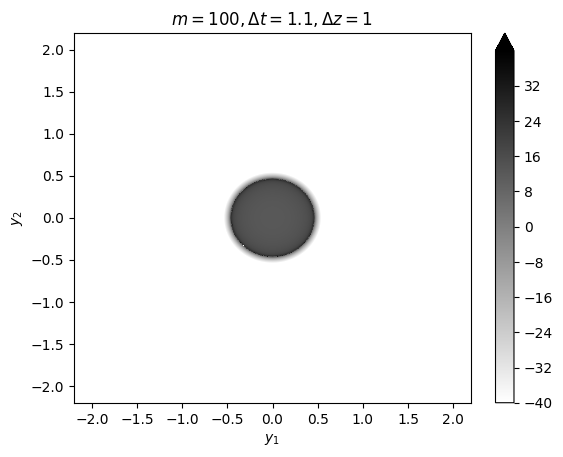} &
    \includegraphics[width=0.3\textwidth]{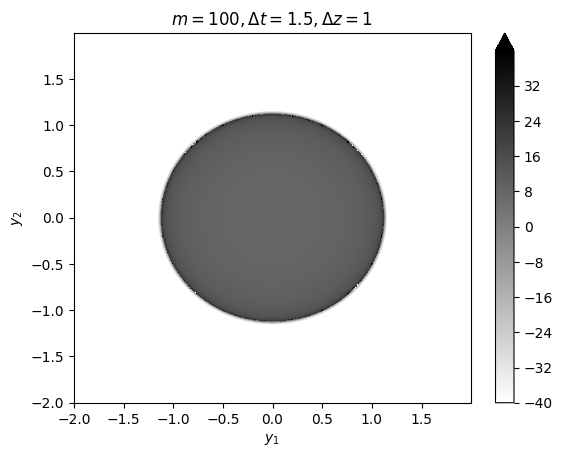} &
    \includegraphics[width=0.3\textwidth]{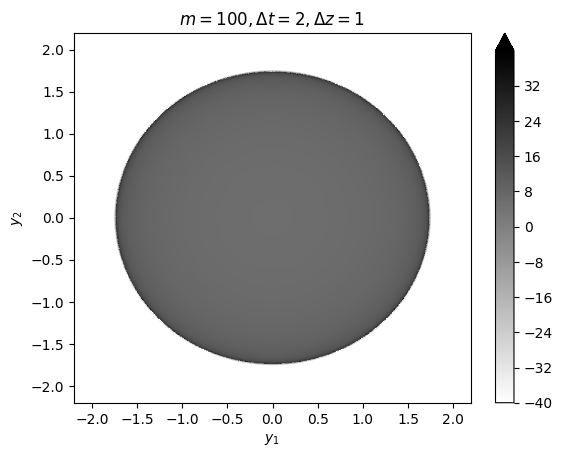}
  \end{tabular}
  \caption{Arrival probability density in the arrival plane for fixed arrival times. The logarithm of the probability density is shown for arrival times $\Delta t=1.1,1.5,2$ (from left to right) and for masses $m=0,10,100$ (from top to bottom). In all cases $\Delta z=1$.}
  \label{fig:arrivalplane}
\end{figure}

It is also instructive to consider the arrival probability density at fixed arrival time $t$, as a function of the location on the arrival plane, i.e., on the target hyperplane at time $t$. This is shown in Figure~\ref{fig:arrivalplane}, on a logarithmic scale. In all cases we set $\Delta z=1$. A particle traveling at the speed of light would hit the target plane at $\Delta t=1$ if travelling without displacement in the $\tilde{y}$-plane. Still supposing a straight path, a particle arriving later with the speed of light would have to arrive with a displacement from the center so that $\Delta x=\sqrt{(\Delta z)^2+\|\Delta\tilde{y}\|^2}=\Delta t$. These particles are indeed the dominant contribution to the probability density and give rise to the black rings in the plots of Figure~\ref{fig:arrivalplane}. The radii of the rings increase with time precisely as expected (from left to right). The areas outside and inside the rings correspond to superluminal and subluminal propagation respectively, assuming straight paths. In the massless case (top row), both are suppressed. With increasing mass, ($m=10$ second row, and $m=100$ third row) the superluminal contribution becomes extremely suppressed, while the subluminal contribution increases significantly. Indeed, propagation at subluminal speeds is exactly what we expect of massive particles.

\subsection{Separating the propagating and evanescent contributions}
\label{sec:seppe}

From a technical and methodological perspective it is interesting to separate the propagating and evanescent contributions to the amplitudes considered. Indeed, the calculations leading to the expressions \eqref{eq:biampl}, \eqref{eq:biampl0}, \eqref{eq:amplr}, \eqref{eq:ampll} were first performed separately for propagating and evanescent sectors, see the appendix, with both being combined afterwards. We consider the individual contributions of the sectors in the following. For simplicity, we consider the bidirectional massless case only.

\begin{figure}
  \centering
  \includegraphics[width=0.5\textwidth]{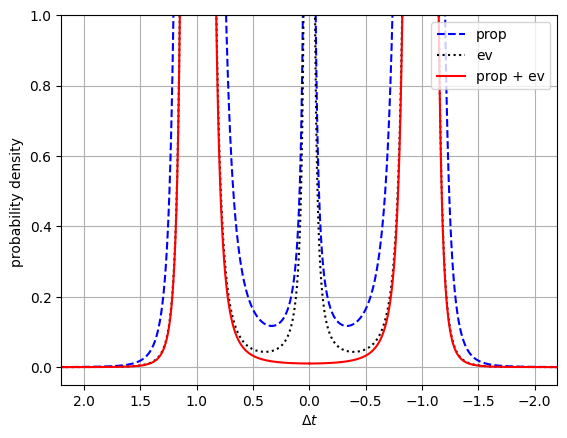}
  \caption{Probability density for the complete amplitude and its propagating and evanescent components as a function of relative incidence time $\Delta t$ for $m=0$, $\Delta z=1$, $\Delta\tilde{y}=(0,0)$.}
  \label{fig:pd_undirected_p_e_c}
\end{figure}

The propagating and evanescent components of the amplitude \eqref{eq:biampl0} of massless hypersurface-localized particles are (see equations \eqref{eq:apR} and \eqref{eq:aeR+L01} of the appendix),
\begin{align}
  \{\Phi_{t,\tilde{y}}^z,\Phi_{t',\tilde{y}'}^{z'}\}^{\Lp}
  & =\frac{1}{2\pi^2}  \frac{1}{\sqrt{\|\Delta\tilde{y}\|^2-(\Delta t)^2}}
     \frac{\left( \sqrt{\|\Delta\tilde{y}\|^2-(\Delta t)^2} +\im\Delta z\right)^2}{\sigma^4},   \label{eq:amplprop}\\
  \{\Phi_{t,\tilde{y}}^z,\Phi_{t',\tilde{y}'}^{z'}\}^{\Le}
  & = \frac{1}{2\pi^2} \frac{1}{\sqrt{\|\Delta\tilde{y}\|^2-(\Delta t)^2}}
      \frac{-\|\Delta\tilde{y}\|^2+(\Delta t)^2 + (\Delta z)^2}{\sigma^4}.
  \label{eq:amplev}
\end{align}
The sum of \eqref{eq:amplprop} and \eqref{eq:amplev} recovers expression \eqref{eq:biampl0}. We can read off a peculiar feature of these amplitudes: In addition to the singularities on the light cone at $\sigma=0$ they exhibit an additional singularity at $(\Delta t)^2=\|\Delta\tilde{y}\|^2$. This is shown in Figure~\ref{fig:pd_undirected_p_e_c}, where the probability densities of the complete amplitude and its components are plotted for $\Delta\tilde{y}=(0,0)$, as in previous plots. The singularity in the components at $\Delta t=0$ corresponds to a simultaneous incidence of the particle on both hypersurfaces, at $x_1=z$ and at $x_1=z'$. This is clearly unphysical. Only when the amplitudes for propagating and evanescent sector are combined, does this spurious singularity cancel. This is further evidence that the twisted quantization prescription as proposed in general form in \cite{CoOe:locgenvac} and more specifically for Klein-Gordon theory on the timelike hypersurface in \cite{CoOe:evanescent} leads to physically sensible results. Conversely, a quantization prescription that is limited to the propagating sector appears to be unphysical. 

A similar spurious singularity appears in the propagating and evanescent components of massive and directed amplitudes as well, although we do not present explicit calculations here.


\section{Conclusions and Outlook}
\label{sec:conclusions}

We have put forward an approach to addressing two fundamental problems in quantum theory: The lack of a time operator and the time-of-arrival problem. To this end we "rotate" the usual spacetime picture of states on an equal-time hyperplane evolving in time (Figure~\ref{fig:tevol}) to a picture of states on a timelike hyperplane "evolving" in space (Figure~\ref{fig:spevol}). Making quantization work in the latter picture is possible due to a recently introduced twisted quantization scheme \cite{CoOe:locgenvac}. More specifically, our present work builds on \cite{CoOe:evanescent}, where both the propagating and evanescent degrees of freedom of the Klein-Gordon field on a timelike hypersurface were quantized consistently. This quantization is reviewed and expanded on in Section~\ref{sec:Hstml}. Compared to \cite{CoOe:evanescent} we improve on the understanding of the evanescent modes and their quantum numbers as follows. We take the hypersurface to be located at a fixed value of the $x_1$-coordinate. In the propagating sector, solutions are combinations of plain waves, and we can distinguish left- and right-moving plain waves, with respect to the $x_1$-position of the hypersurface, directly from the motion of the waveforms. Upon quantization, this leads to a corresponding binary quantum number. In the evanescent sector, solutions have exponential behavior in the $x_1$-direction and there seems to be no way to read off from the wave forms a distinction into left- and right-moving solutions. However, a similar situation occurs for the timelike hypercylinder, with modes flowing into or out of the interior \cite{Oe:quanthcyl}. In that case, measuring the energy flow through the hypercylinder via the energy-momentum tensor allowed to extend a distinction between inflowing and outflowing solutions to the evanescent sector \cite{CoOeZa:udwevanescent}. In the present case, we show that we can similarly use the flow of energy through the hypersurface to decompose evanescent solutions into left- and right-moving components. In contrast to the hypercylinder case, there is an arbitrariness to this decomposition, and any such decomposition is not conserved under translations. However, the total flux is conserved under translations, which allows us to consistently work with one specific decomposition by fixing a reference hypersurface. The corresponding binary quantum number is essential in Section~\ref{sec:unidir} to distinguish left- and right-moving particles in the time-of-arrival problem.

In Section~\ref{sec:spaceop} we review the Newton-Wigner localized one-particle states and operator with an emphasis on the corresponding Positive Operator Valued Measure (POVM), essential for a consistent probability interpretation in the measurement of particle positions. The probability of particle detection as a function of position is reviewed in Section~\ref{sec:positionprob}. In the massless case, the probability is peaked on the light-cone, while increasing mass leads to increasing probability for subluminal propagation (see Figure~\ref{fig:pdenstint}). As is well known, there is also a non-vanishing, albeit exponentially suppressed probability for apparent superluminal propagation. The consensus seems to be that this is not due to actual superluminal particle motion, but to an uncertainty inherent to any particle localization scheme in quantum theory. Superluminal motion and its relationship to localization in quantum theory and the Newton-Wigner position operator has been addressed in the literature, see e.g.\ \cite{Ruijsenaars:1981sm} as well as \cite{Hegerfeldt:1985fy} and references therein.

We construct in Section~\ref{sec:timeop} one-particle states localized in time and two spatial directions on a timelike hyperplane. In close analogy to the Newton-Wigner setting, these give rise to a POVM. In spite of the preparative work of Section~\ref{sec:Hstml} this construction is not quite as straightforward as the Newton-Wigner one. This is because in addition to the propagating sector, there appears now an evanescent sector in the solution space and the latter is subject to different quantization rules. In particular, one-particle states are no longer labeled by real phase space elements, but by certain complexified ones. This requires the use of an identification map between the two. A guiding principle for the construction remains a delta-function normalization and completeness relation of particle states (see equations \eqref{eq:tdeltanorm} and \eqref{eq:tcompl}), as in the Newton-Wigner setting. The resulting POVM permits a consistent probability interpretation in spite of the fact that the spatial "evolution" between timelike hyperplanes turns out to be non-unitary \cite{CoOe:evanescent}. It also yields the \emph{time operator} which measures the incidence time of a particle on the timelike hypersurface.

We address the \emph{time-of-arrival problem} in Section~\ref{sec:toa}, armed with the time-localized states constructed previously. As a first step, and in analogy to the Newton-Wigner setting, we consider transitions between time-localized states on spatially separated timelike hyperplanes (Section~\ref{sec:bincprob}). Given a particle incident on the (say) left-hand hyperplane at a given time, the probability for an incidence on the right-hand hyperplane peaks at two times, one before the incidence on the left-hand hyperplane, and one after (see Figure~\ref{fig:timedetection}). The later peak is due to the particle moving from the left-hand to the right-hand hyperplane, while the earlier peak is due to the particle moving from the right-hand to the left-hand hyperplane. We call this setting the \emph{bidirectional incidence problem}, i.e., the problem of determining the probability for the incidence of a particle on the second hyperplane as a function of time (and place), given an incidence on the first hyperplane at a specific time and place. The expression for the amplitude and thus also probability (density) is almost identical to the Newton-Wigner one, except for the exchange between a spatial and temporal distance, compare expressions \eqref{eq:propampleval} and \eqref{eq:biampl}. Indeed, the plots in the massless case look almost the same in spite of the exchange of a spatial and a temporal direction, compare Figures~\ref{fig:posdetection} and \ref{fig:timedetection}. We have an exponential decay of the probability density away from the light-cone. In particular, there is an exponentially suppressed superluminal contribution in both settings. That is, the time-localization of the constructed states is subject to the same fundamental uncertainty as the spatial localization in the Newton-Wigner setting.

In order to proceed to the actual time-of-arrival problem (Section~\ref{sec:unidir}), we have to deal with the fact that the time-localized states carry no intrinsic information about whether they are left- or right-moving. They are undetermined in this respect. This implies that in a measurement context we cannot a priori distinguish between emission and absorption events, as evidenced in the bidirectional incidence problem. This is a crucial difference to the Newton-Wigner setting. There, particles can only move in one direction perpendicular to the hypersurface, namely to the future. Using the decomposition into left- and right-moving sectors introduced in Section~\ref{sec:Hstml}, the time-localized one-particle states can correspondingly be decomposed into left- and right-moving one-particle states. Replacing a time-localized particle state on the left-hand hypersurface by a time-localized particle state that in addition is purely right-moving, leads to a probability distribution for the incidence time on the right-hand hypersurface that is peaked at a single time. The early time peak is gone, only the late time peak appears, compare Figure~\ref{fig:timedetection}, black line. This is because we can now be certain that the particle moves to the right which implies that it passes the left-hand hypersurface earlier than the right-hand hypersurface. This is the proper time-of-arrival problem. As we have mentioned already, in the massless case the probability density peaks at propagation with the speed of light, decaying exponentially for subluminal and superluminal propagation. With increasing mass, the probability for subluminal propagation increases substantially, while the probability for superluminal propagation decreases even more strongly, see Figure~\ref{fig:pdens-arrival}. We also consider the probability distribution for arrival in the $x_2,x_3$-plane given that the arrival happens at a fixed time, see Figure~\ref{fig:arrivalplane}, confirming this same behavior, in accordance with physical expectations.

While our results paint a consistent and, we think, convincing picture of the merits of our approach to time-localized states, a time operator, and the time-of-arrival problem, we include in Section~\ref{sec:seppe} a small technical exercise to underline the importance of including both, propagating and evanescent sectors in the quantization scheme used. In particular, we show that restricting to either the propagating or the evanescent sector only, leads to an unphysical singularity in the bidirectional incidence problem, corresponding to instantaneous particle propagation. In general, our results can also be seen to provide additional support for the "correctness" of the novel twisted quantization scheme introduced in \cite{CoOe:locgenvac}. Previous important tests have been the results of \cite{CoOe:evanescent,Oe:quanthcyl,CoOeZa:udwevanescent}.

One of the first questions one might ask with respect to the present work is how it compares to previous works on the time operator and on the time-of-arrival problem. With respect to the time operator it appears difficult to establish a meaningful comparison to any operator defined on an instantaneous state space, due to the underlying conceptual change involved in replacing a spacelike with a timelike hypersurface. As for the time-of-arrival problem, what we want to predict in the end are probability distributions for arrival times of a particle (such as shown in Figure~\ref{fig:pdens-arrival}). These should definitely be comparable in principle, not only between different proposals, but eventually also with experiments. One challenge here lies in characterizing properties of the particle, additional to its initial location and emission time, such as emission direction, momentum, and energy. In a quantum theory, not all these parameters can have precise values at the same time. In particular, the states we have defined in this work have a precise incidence time and initial location, but undetermined energy and momenta (except possibly for binary information about the propagation direction). We are not aware of any other approach to the time-of-arrival problem that involves states with these characteristics, impeding for the moment a quantitative comparison.
However, uncertainty in energy can be traded for uncertainty in time and uncertainty in momentum for uncertainty in position. This suggests introducing a notion of Gaussian state to balance different particle properties and their uncertainties. This would be an important step for future development, but is beyond the scope of the present work.

We have limited ourselves in the present work to the Klein-Gordon field, due to its simplicity. This might be useful as a crude model for simple particles, where it is reasonable to neglect spin, charge, internal structure etc. However, for a realistic description of photons we would need to deal with the electromagnetic field. To this end, the twisted quantization scheme of \cite{CoOe:locgenvac} would have to be adapted to the electromagnetic field, which appears quite feasible, and indeed desirable for a number of applications. Then, the actual quantization on a timelike hyperplane has to be carried out, in analogy to \cite{CoOe:evanescent}, but that would appear relatively straightforward. For the case of fermionic particles, such as electrons, it is less clear whether the twisted quantization scheme can be carried over, or a different approach is needed. These are clearly important questions that should be addressed in future research.

There are other interesting applications for time-localized states than the time-of-arrival problem. A closely related problem is that of tunneling time, see \cite{Hauge:1989zz,Landauer:1994zz,Winful:20061}. In this case, a particle passes a classically forbidden region, e.g.\ due to a potential barrier, and one is interested in the time the particle takes in crossing the barrier compared to the time without a barrier. For sufficiently simple potentials this would be easy to implement with the methods at hand.

We have already alluded to difficulties and issues with the Newton-Wigner localization scheme that are inherited by our approach, such as apparent superluminal propagation. However, the main principle underlying our approach, summarized as "rotating by 90° in spacetime", may very well be applicable to other spatial localization schemes, that do not suffer from these issues. Again, this would be material for future work.

\subsection*{Acknowledgments}

This work was partially supported by UNAM-PAPIIT project grant IN106422 and UNAM PASPA-DGAPA.

\appendix



\section{Amplitude Calculations}
\label{sec:app}

This appendix provides some details on the calculation of the amplitudes discussed in Section~\ref{sec:toa}.

\subsection{Propagating particles}

The amplitude of propagating right-moving particles is constructed in terms of the modes \eqref{eq:modeR} and \eqref{eq:modeRbar} as
\begin{equation}
    \{\Phi_{\Lr,t,\tilde{y}}^z,\Phi_{\Lr,t',\tilde{y}'}^{z'}\}^{\Lp}
    = \int_{E>E_\parallel}
    \frac{\xd^2\tilde{k}\,\xd E}{(2\pi)^3}\,
    e^{-\im (E(t'-t)-\tilde{k}(\tilde{y}'-\tilde{y}))} e^{\im k_1 (z'-z)}; \label{eq:arp}
\end{equation}
 the two-dimensional integral over $\tilde{k}$ can be rewritten using expression 10.9.2 of \cite{NIST:DLMF} in the form
\begin{equation}
    \{\Phi_{\Lr,t,\tilde{y}}^z,\Phi_{\Lr,t',\tilde{y}'}^{z'}\}^{\Lp}
    = \int_0^{\infty} \frac{\xd \tilde{k}}{(2 \pi)^2 }  \tilde{k} J_0 \left( \tilde{k} |\tilde{y}'-\tilde{y}|\right)
    \int_{E_\parallel}^{\infty} \xd E \, e^{- E (\epsilon+\im(t'-t))  +\im k_1(z'-z)},
\end{equation}
where expression 10.9.2 of \cite{NIST:DLMF} has been used, and a small imaginary part has been added to $(t'-t)$ in order to guarantee convergence. Then by changing the integration variable in the inner integral (from $E$ to $k_1$) and integrating by parts leading to 
\begin{align}
   & \{\Phi_{\Lr,t,\tilde{y}}^z,\Phi_{\Lr,t',\tilde{y}'}^{z'}\}^{\Lp}
    =\frac{1}{ (\epsilon+\im (t'-t))(2\pi)^2}  \nonumber\\
    &\int_{0}^{\infty} \xd k_1 \, k_1  e^{\im k_1(z'-z)}  \left[    e^{- \sqrt{k_1^2+m^2} (\epsilon+\im (t'-t)) }   - |\tilde{y}'-\tilde{y}| 
    \int_0^{\infty} \xd \tilde{k} \, J_1 \left( \tilde{k} |\tilde{y}'-\tilde{y}|\right) 
    e^{- \sqrt{k_1^2+ \tilde{k}^2+m^2} (\epsilon+\im (t'-t)) }\right].
\end{align}
The integral over $\tilde{k}$ can be evaluated with expression 6.637.1 of \cite{GR:tables} according to the following identifications: $\gamma = |\tilde{y}'-\tilde{y}|$, $\nu=1$, $\alpha = (\epsilon+\im (t'-t))$ and $\beta^2 = k_1^2+m^2$, 
\begin{equation}
  \{\Phi_{\Lr,t,\tilde{y}}^z,\Phi_{\Lr,t',\tilde{y}'}^{z'}\}^{\Lp} =\int_{0}^{\infty} \xd k_1 \, k_1  e^{\im k_1(z'-z)} \frac{1}{ (2\pi)^2}     \frac{ e^{-\sqrt{k_1^2+m^2} \sqrt{(\epsilon+\im (t'-t))^2 +  |\tilde{y}'-\tilde{y}|^2}}}{  \sqrt{(\epsilon+\im (t'-t))^2+ |\tilde{y}'-\tilde{y}|^2}}.
\end{equation}
In order to perform the integral over $k_1$ we consider the massless case,
\begin{equation}
   \{\Phi_{\Lr,t,\tilde{y}}^z,\Phi_{\Lr,t',\tilde{y}'}^{z'}\}^{\Lp}
   = \frac{1}{ (2\pi)^2} \frac{1}{ \sqrt{(\epsilon+\im(t'-t))^2+ |\tilde{y}'-\tilde{y}|^2}}
     \frac{1}{ \left(\im (z'-z) - \sqrt{(\epsilon+\im(t'-t))^2+ |\tilde{y}'-\tilde{y}|^2}\right)^2}.
      \label{eq:apR}
\end{equation}
The amplitude of propagating left-moving particles, namely
\begin{equation}
    \{\Phi_{\Ll,t,\tilde{y}}^z,\Phi_{\Ll,t',\tilde{y}'}^{z'}\}^{\Lp}
     = \int_{E>E_\parallel}
    \frac{\xd^2\tilde{k}\,\xd E}{(2\pi)^3}\,
    e^{\im (E(t'-t)-\tilde{k}(\tilde{y}'-\tilde{y}))} e^{\im k_1 (z'-z)}, 
\end{equation}
 can be obtained from \eqref{eq:apR} by the interchange $t' \leftrightarrow t$. Therefore, the sum of the amplitudes of propagating left- and right-moving particles turns out to be
\begin{multline}
     \{\Phi_{\Lr,t,\tilde{y}}^z,\Phi_{\Lr,t',\tilde{y}'}^{z'}\}^{\Lp}+\{\Phi_{\Ll,t,\tilde{y}}^z,\Phi_{\Ll,t',\tilde{y}'}^{z'}\}^{\Lp} \\
     =\frac{1}{2\pi^2}  \frac{1}{ \sqrt{-(t'-t)^2+|\tilde{y}'-\tilde{y}|^2} }
      \frac{1}{ \left( \sqrt{-(t'-t)^2+|\tilde{y}'-\tilde{y}|^2} -\im(z'-z)\right)^2}.
      \label{eq:apR+L}
\end{multline}

\subsection{Evanescent particles}
The modes (\ref{eq:modeRev}-\ref{eq:modeLbarev}) lead to the amplitude of evanescent right-moving particles in the form
\begin{equation}
    \{\Phi_{\Lr,t,\tilde{y}}^z,\Phi_{\Lr,t',\tilde{y}'}^{z'}\}^{\Le}
     = \int_{E<E_\parallel}
    \frac{\xd^2\tilde{k}\,\xd E}{(2\pi)^3}\,
    e^{-\im (E(t'-t)-\tilde{k}(\tilde{y}'-\tilde{y}))} e^{-k_1 (z'-z)},\label{eq:are} 
\end{equation}
which can be written as
\begin{equation}
    \{\Phi_{\Lr,t,\tilde{y}}^z,\Phi_{\Lr,t',\tilde{y}'}^{z'}\}^{\Le}
    = \int_0^{\infty} \frac{\xd \tilde{k}}{(2 \pi)^2 }  \tilde{k} J_0 \left( \tilde{k} |\tilde{y}'-\tilde{y}|\right)
\int_{0}^{\frac{\pi}{2}} \xd \theta \, E_\parallel \cos \theta \, e^{-\im E_\parallel (t'-t) \sin\theta  - E_\parallel (z'-z) \cos \theta}
\end{equation}
where the variable $\theta$ is related to the energy $E$ as $E=E_\parallel \sin \theta$. In the massless case ($E_\parallel = \tilde{k}$), the integral over $\tilde{k}$ may be evaluated using expression 6.611.1 of \cite{GR:tables}, with the following identifications: $\nu=0$, $b= |\tilde{y}'-\tilde{y}|$ and $\alpha = \left[\im (t'-t) \sin\theta  +  (z'-z) \cos \theta \right]$,
\begin{equation}
    \{\Phi_{\Lr,t,\tilde{y}}^z,\Phi_{\Lr,t',\tilde{y}'}^{z'}\}^{\Le} 
    =\frac{\partial}{\partial (z'-z)} \frac{1}{(2\pi)^2}
    \left[
    \frac{\im (t'-t)}{\sigma^2 \sqrt{(z'-z)^2 + |\tilde{y}'-\tilde{y}|^2 }}
    +\frac{(z'-z)}{\sigma^2 \sqrt{(\im(t'-t))^2 +  |\tilde{y}'-\tilde{y}|^2}}
    \right] .
\end{equation}
As in the propagating case, the amplitude of evanescent left moving particles, namely
\begin{equation}
    \{\Phi_{\Ll,t,\tilde{y}}^z,\Phi_{\Ll,t',\tilde{y}'}^{z'}\}^{\Le}
     = \int_{E<E_\parallel}
    \frac{\xd^2\tilde{k}\,\xd E}{(2\pi)^3}\,
    e^{\im (E(t'-t)-\tilde{k}(\tilde{y}'-\tilde{y}))} e^{-k_1 (z'-z)},
\end{equation}
 can be obtained from \eqref{eq:apR} by the interchange $t' \leftrightarrow t$; the sum of the amplitudes of evanescent left- and right-moving particles is then equal to
\begin{multline}
  \{\Phi_{\Lr,t,\tilde{y}}^z,\Phi_{\Lr,t',\tilde{y}'}^{z'}\}^{\Le} +  \{\Phi_{\Ll,t,\tilde{y}}^z,\Phi_{\Ll,t',\tilde{y}'}^{z'}\}^{\Le} \\
  = \frac{1}{2\pi^2} \frac{1}{\sqrt{(\im(t'-t))^2 +  |\tilde{y}'-\tilde{y}|^2}}
  \frac{(t'-t)^2 + (z'-z)^2- |\tilde{y}'-\tilde{y}|^2}{\sigma^4}.
  \label{eq:aeR+L01}
\end{multline}

\subsection{Sum of propagating and evanescent amplitudes}

With the results presented above we can separate the contribution of right-moving particles, both propagating and evanescent, from the left-moving ones. In particular, the sum of the amplitudes for propagating and evanescent right-moving particles results to be
\begin{multline}
\{\Phi_{\Lr,t,\tilde{y}}^z,\Phi_{\Lr,t',\tilde{y}'}^{z'}\}^{\Lp} +  \{\Phi_{\Lr,t,\tilde{y}}^z,\Phi_{\Lr,t',\tilde{y}'}^{z'}\}^{\Le} \\
= \frac{\im (z'-z)   \left[ 2 \left((z'-z)^2+|\tilde{y}'-\tilde{y}|^2\right)^{1/2}- (t'-t)+ \im \epsilon \right]  }{(2\pi)^2   \left[  \left((z'-z)^2+|\tilde{y}'-\tilde{y}|^2\right)^{1/2} - (t'-t) + \im \epsilon \right]^2  \left((z'-z)^2+|\tilde{y}'-\tilde{y}|^2\right)^{3/2} }.
\label{eq:leftampl}
\end{multline}
For left-moving particles we have
\begin{multline}
\{\Phi_{\Ll,t,\tilde{y}}^z,\Phi_{\Ll,t',\tilde{y}'}^{z'}\}^{\Lp} +  \{\Phi_{\Ll,t,\tilde{y}}^z,\Phi_{\Ll,t',\tilde{y}'}^{z'}\}^{\Le}  \\
= \frac{\im (z'-z)   \left[ 2 \left((z'-z)^2+|\tilde{y}'-\tilde{y}|^2\right)^{1/2}+ (t'-t)+ \im \epsilon \right]  }{(2\pi)^2   \left[  \left((z'-z)^2+|\tilde{y}'-\tilde{y}|^2\right)^{1/2} + (t'-t) + \im \epsilon \right]^2  \left((z'-z)^2+|\tilde{y}'-\tilde{y}|^2\right)^{3/2} }.
\label{eq:rightampl}
\end{multline}
Finally, the amplitude resulting from all the contributions for massless particles is
\begin{equation}
\{\Phi_{t,\tilde{y}}^z,\Phi_{t',\tilde{y}'}^{z'}\}^{\Lp} +  \{\Phi_{t,\tilde{y}}^z,\Phi_{t',\tilde{y}'}^{z'}\}^{\Le} =  \frac{\im (z'-z)}{\pi^2 \sigma^4 }.
\label{eq:ampl-massless}
\end{equation}

For massive particles, the sum of all contributions, propagating and evanescent for right- and left-moving particles is
\begin{align}
\{\Phi_{t,\tilde{y}}^z,\Phi_{t',\tilde{y}'}^{z'}\}^{\Lp} +  \{\Phi_{t,\tilde{y}}^z,\Phi_{t',\tilde{y}'}^{z'}\}^{\Le} =
 \int_{0}^{\infty}
    \frac{\xd^2\tilde{k}\,\xd E}{4\pi^3}\, \cos \left(E(t'-t) \right)
    e^{-\im \tilde{k}(\tilde{y}'-\tilde{y})} e^{\im k_1 (z'-z)},
\end{align}
where the integral over the energy $E$ can be evaluated in terms of expression 3.961.2 of \cite{GR:tables}, leading to
\begin{multline}
\{\Phi_{t,\tilde{y}}^z,\Phi_{t',\tilde{y}'}^{z'}\}^{\Lp} +  \{\Phi_{t,\tilde{y}}^z,\Phi_{t',\tilde{y}'}^{z'}\}^{\Le}
\\ = - \frac{\im}{\pi}  \frac{\partial}{\partial (z'-z)} 
\int_{0}^{\infty} \frac{\xd k}{2\pi} \, k  J_0( k|\tilde{y}-\tilde{y}'|)  K_0 \left(  E_\parallel \sqrt{(z'-z)^2 - (t'-t)^2}\right)
\end{multline}
It turns out to be convenient to distinguish between real and imaginary values of the square root appearing in the argument of the modified Bessel function. In particular,
\begin{enumerate}
	\item for $\sigma^2 >0$, we have
	\begin{multline}
	\{\Phi_{t,\tilde{y}}^z,\Phi_{t',\tilde{y}'}^{z'}\}^{\Lp} +  \{\Phi_{t,\tilde{y}}^z,\Phi_{t',\tilde{y}'}^{z'}\}^{\Le}	\\
	=  \frac{1}{2}  
	\int_{0}^{\infty} \frac{\xd k}{2\pi} \, k  J_0( k|\tilde{y}-\tilde{y}'|)  H_1^{(2)} \left(   E_\parallel \sqrt{(t'-t)^2 - (z'-z)^2}\right)  \frac{E_\parallel(z'-z)}{\sqrt{(t'-t)^2 - (z'-z)^2}}
	\end{multline}
	The integral can be evaluated with expression 6.596.9 of \cite{GR:tables}, 
	\begin{equation}
	\{\Phi_{t,\tilde{y}}^z,\Phi_{t',\tilde{y}'}^{z'}\}^{\Lp} +  \{\Phi_{t,\tilde{y}}^z,\Phi_{t',\tilde{y}'}^{z'}\}^{\Le}
	=  \frac{m^2}{4 \pi}  \frac{ z'-z}{ \sigma^2} H_{2}^{(2	)} \left( m \sqrt{\sigma^2}\right).
	\label{eq:ampl-tm}
	\end{equation}
	where we have used the identity 10.4.2 of \cite{NIST:DLMF}. 

	\item For $\sigma^2<0$, we obtain
	\begin{equation}
	\{\Phi_{t,\tilde{y}}^z,\Phi_{t',\tilde{y}'}^{z'}\}^{\Lp} +  \{\Phi_{t,\tilde{y}}^z,\Phi_{t',\tilde{y}'}^{z'}\}^{\Le}
	=  - \im \frac{m^2}{2\pi^2}  \frac{z'-z}{\sigma^2 } K_2 \left(m \sqrt{-\sigma^2}\right).
	\label{eq:ampl-sp}
	\end{equation}
\end{enumerate}
Notice that the limit $m \to 0$ of both amplitudes \eqref{eq:ampl-tm} and \eqref{eq:ampl-sp} coincides with the expression of the massless amplitude \eqref{eq:ampl-massless}.

\newcommand{\eprint}[1]{\href{https://arxiv.org/abs/#1}{#1}}
\bibliographystyle{stdnodoi} 
\bibliography{stdrefsb}

\begin{thebibliography}{10}
\providecommand{\url}[1]{\texttt{#1}}
\providecommand{\urlprefix}{URL }
\providecommand{\selectlanguage}[1]{\relax}
\providecommand{\eprint}[2][]{\url{#2}}

\bibitem{Flu:encphys5-1}
S.~Flügge (ed.), \textit{Encyclopedia of Physics}, vol. 5/1, Springer, Berlin, 1958.

\bibitem{Allcock:1969cq}
G.~R. Allcock, \textit{The time of arrival in quantum mechanics}, Annals Phys. \textbf{53} (1969) I 253--285, II 286--310, III 311--348.

\bibitem{MuLe:arrivaltime}
J.~Muga, C.~Leavens, \textit{Arrival time in quantum mechanics}, Physics Reports \textbf{338} (2000) 353--438.

\bibitem{Oe:dmf}
R.~Oeckl, \textit{A Positive Formalism for Quantum Theory in the General Boundary Formulation}, Found. Phys. \textbf{43} (2013) 1206--1232, \eprint{1212.5571}.

\bibitem{Oe:posfound}
R.~Oeckl, \textit{A local and operational framework for the foundations of physics}, Adv. Theor. Math. Phys. \textbf{23} (2019) 437--592, \eprint{1610.09052v3}.

\bibitem{Oe:boundary}
R.~Oeckl, \textit{A ``general boundary'' formulation for quantum mechanics and quantum gravity}, Phys. Lett. \textbf{B 575} (2003) 318--324, \eprint{hep-th/0306025}.

\bibitem{Oe:timelike}
R.~Oeckl, \textit{States on timelike hypersurfaces in quantum field theory}, Phys. Lett. \textbf{B 622} (2005) 172--177, \eprint{hep-th/0505267}.

\bibitem{Oe:kgtl}
R.~Oeckl, \textit{General boundary quantum field theory: Timelike hypersurfaces in Klein-Gordon theory}, Phys. Rev. \textbf{D 73} (2006) 065017, \eprint{hep-th/0509123}.

\bibitem{CoOe:spsmatrix}
D.~Colosi, R.~Oeckl, \textit{S-matrix at spatial infinity}, Phys. Lett. \textbf{B 665} (2008) 310--313, \eprint{0710.5203}.

\bibitem{CoOe:smatrixgbf}
D.~Colosi, R.~Oeckl, \textit{Spatially asymptotic S-matrix from general boundary formulation}, Phys. Rev. \textbf{D 78} (2008) 025020, \eprint{0802.2274}.

\bibitem{CoOe:locgenvac}
D.~Colosi, R.~Oeckl, \textit{Locality and General Vacua in Quantum Field Theory}, SIGMA \textbf{17} (2021) 073, 83 pages, \eprint{2009.12342}.

\bibitem{CoOe:evanescent}
D.~Colosi, R.~Oeckl, \textit{Evanescent Particles}, Int. J. Mod. Phys. A \textbf{36} (2021) 2150194, 23 pages, \eprint{2104.12321}.

\bibitem{Grot:1996xu}
N.~Grot, C.~Rovelli, R.~S. Tate, \textit{Time-of-arrival in quantum mechanics}, Phys. Rev. A \textbf{54} (1996) 4676--4690, \eprint{quant-ph/9603021}.

\bibitem{AnSa:time_2019}
C.~Anastopoulos, N.~Savvidou, \textit{Time of arrival and localization of relativistic particles}, J. Math. Phys. \textbf{60} (2019) 032301.

\bibitem{Kijowski:1974jx}
J.~Kijowski, \textit{On the Time Operator in Quantum Mechanics and the Heisenberg Uncertainty Relation for Energy and Time}, Rept. Math. Phys. \textbf{6} (1974) 361--386.

\bibitem{Woo:geomquant}
N.~M.~J. Woodhouse, \textit{Geometric Quantization}, 2nd ed., Oxford University Press, Oxford, 1991.

\bibitem{BiDa:qftcurved}
N.~D. Birrell, P.~C.~W. Davies, \textit{Quantum Fields in Curved Space}, Cambridge University Press, Cambridge, 1982.

\bibitem{NeWi:locstates}
T.~D. Newton, E.~P. Wigner, \textit{Localized States for Elementary Systems}, Rev. Mod. Phys. \textbf{21} (1949) 400--406.

\bibitem{Wig:localizability}
A.~S. Wightman, \textit{On the Localizability of Quantum Mechanical Systems}, Rev. Mod. Phys. \textbf{34} (1962) 845--872.

\bibitem{CoOe:vaclag}
D.~Colosi, R.~Oeckl, \textit{The Vacuum as a Lagrangian subspace}, Phys. Rev. \textbf{D 100} (2019) 045018, \eprint{1903.08250}.

\bibitem{Oe:quanthcyl}
R.~Oeckl, \textit{Scattering of Evanescent Particles}, Int. J. Mod. Phys. A \textbf{38} (2023) 2350081, 25 pages, \eprint{2105.07600}.

\bibitem{CoOeZa:udwevanescent}
D.~Colosi, R.~Oeckl, A.~Zampeli, \textit{Interaction of evanescent particles with an Unruh-DeWitt detector}, Phys. Rev. \textbf{D 109} (2024) 025009, 27 pages, \eprint{2310.06716}.

\bibitem{Ruijsenaars:1981sm}
S.~N.~M. Ruijsenaars, \textit{On Newton-Wigner Localization and Superluminal Propagation Speeds}, Annals Phys. \textbf{137} (1981) 33--43.

\bibitem{Hegerfeldt:1985fy}
G.~C. Hegerfeldt, \textit{Violation of causality in relativistic quantum theory?}, Phys. Rev. Lett. \textbf{54} (1985) 2395--2398.

\bibitem{Hauge:1989zz}
E.~H. Hauge, J.~A. Stovneng, \textit{Tunneling times: a critical review}, Rev. Mod. Phys. \textbf{61} (1989) 917--936.

\bibitem{Landauer:1994zz}
R.~Landauer, T.~Martin, \textit{Barrier interaction time in tunneling}, Rev. Mod. Phys. \textbf{66} (1994) 217--228.

\bibitem{Winful:20061}
H.~G. Winful, \textit{Tunneling time, the Hartman effect, and superluminality: A proposed resolution of an old paradox}, Physics Reports \textbf{436} (2006) 1--69.

\bibitem{NIST:DLMF}
\textit{NIST Digital Library of Mathematical Functions}, Release 1.0.19 of 2018-06-22, \urlprefix\url{http://dlmf.nist.gov/}.

\bibitem{GR:tables}
I.~S. Gradshteyn, I.~M. Ryzhik, \textit{Tables of integrals, series, and products}, Academic Press, New York, 1980.

\end{thebibliography}
\end{document}